\documentclass[a4paper,12pt]{article}
\usepackage{epsfig}
\date{\today}
\def\lsim{\raise0.3ex\hbox{$\;<$\kern-0.75em\raise-1.1ex\hbox{$\sim\;$}}}
\def\gsim{\raise0.3ex\hbox{$\;>$\kern-0.75em\raise-1.1ex\hbox{$\sim\;$}}}

\usepackage{graphics}

\newcommand{\da}[1]{16\pi^2\;\frac{d #1}{dt}}
\newcommand{\be}{\begin{eqnarray}}
\newcommand{\ee}{\end{eqnarray}}

\newcommand{\n}{\nonumber\\}
\newcommand{\nn}{\tilde{\nu}}

\def\bea{\begin{eqnarray}}
\def\eea{\end{eqnarray}}

\usepackage{color}
\usepackage{epsfig}
 \normalsize

\begin{document}
\begin{center}
{\large \bf The $B-L$ Supersymmetric Standard Model\\[0.25cm]
 with Inverse Seesaw at the Large Hadron Collider }
\end{center}
\vspace{.3cm}
\begin{center}
S. Khalil$^{1}$ and  S. Moretti$^{2}$ \\
\vspace{.3cm}
{\small \it $^{1}$ Center for Fundamental Physics, Zewail City for Science and Technology, 6 October City, Cairo, Egypt.}\\
{\small \it $^{3}$ School of Physics \& Astronomy, University of Southampton, Highfield, Southampton, UK.}
\end{center}
\vspace{.3cm}
\vskip 0.3cm
\begin{center}
\small{\bf Abstract}\\[3mm]

\begin{minipage}[h]{14.5cm}
We review the TeV scale $B-L$ extension of the Minimal Supersymmetric Standard Model (BLSSM) where an Inverse Seesaw mechanism
of light neutrino mass generation is naturally implemented and concentrate on its hallmark manifestations at the Large Hadron Collider (LHC).

\end{minipage}
\end{center}
\vskip 0.3cm
\section{\large{\bf Introduction}}
The solid experimental evidence for neutrino oscillations, pointing towards non-vanishing neutrino masses, is one
of the few firm hints for physics Beyond the Standard Model (BSM).
Neutrinos are strictly massless in the SM due to two main reasons: $(i)$ the absence of right-handed
neutrinos; $(ii)$ the SM has an exact global Baryon minus Lepton $(B-L)$ number conservation. However, the
 minimal extension of the SM, based on the gauge group $SU(3)_C \times SU(2)_L
\times U(1)_Y \times U(1)_{B-L}$, can account for light neutrino masses through either a Type-I
Seesaw or an  Inverse Seesaw (IS) mechanism \cite{Khalil:2006yi,Khalil:2010iu}\footnote{For old attempts at analysing a high scale
$B-L$ extension of the SM, see Ref. \cite{B-L}.}.

In the type-I Seesaw mechanism
right-handed neutrinos acquire Majorana masses at the $B-L$ symmetry breaking scale, which can be
related to the Supersymmetry (SUSY) breaking scale, {i.e.}, ${\cal O}(1)$ TeV, therefore the Yukawa neutrino coupling must be $\lsim {\cal O}(10^{-6})$ \cite{Khalil:2007dr}. In contrast, in the IS case, these Majorana masses are not allowed by the $B-L$ gauge symmetry and another pair of SM gauge singlet
fermions with tiny masses $({\cal O}(1)$ keV) must be introduced. In this case, there is no severe constraint imposed on the neutrino Yukawa couplings and the possibility of testing this type of model at the LHC is quite feasible.  Moreover, in the limit of the above mentioned tiny mass,  we have essentially massless light neutrinos. Therefore, such a small
scale can be considered as a slight breaking of a global symmetry, hence,  according to 't Hooft criteria, the smallness of this scale becomes natural.
One of these two singlet fermions  couple to right-handed neutrinos and is involved in generating the light neutrino masses. The other
singlet (which is usually called inert or sterile neutrino) is completely decoupled and interacts only
through the $B-L$ gauge boson, therefore it may account for warm Dark Matter (DM) \cite{El-Zant:2013nta}, see also Refs. \cite{Basso:2012gz,Khalil:2006yi5}.

In both scenarios, this $B-L$ model induces several testable signals at the LHC involving the new predicted
particles: a $Z'$ (neutral gauge boson associated with the $U(1)_{B-L}$ group), an extra Higgs state (an additional singlet
state is introduced to break the gauge group $U(1)_{B-L}$ spontaneously) and three (Type-I) or six (IS)
heavy neutrinos, $\nu_h$ (that are required to cancel the associated anomaly and are necessary for the
consistency of the model). This is the setup for the non-SUSY sector of the $B-L$ scenario, which is well established
in the literature (see Refs.~\cite{Khalil:2012gs,Khalil:2013in} for a review of its main phenomenological manifestations).

It is the purpose of this paper to review its Supersymmetric version, the BLSSM, particularly in the IS framework.
The reason to concentrate on the IS version of the BLSSM is twofold. On the one hand, its phenomenological exploitation is more
recent and less evolved with respect to the type-I Seesaw mechanism case, hence the need of reviewing its status in view of the
ongoing Run 2 stage of the LHC. Secondly, the former appears to have some intriguing and testable features in the Higgs
sector that the latter has not (as we shall try to emphasise), again, calling for a timely assessment of the LHC sensitivity to it, given the high priority that Higgs physics has for Run 2.

The plan of the paper is as follows. In the next section we proceed to the construction of the BLSSM Lagrangian.
In Sect.~\ref{sect:radiative} we describe how dynamical Electro-Weak Symmetry Breaking (EWSB) occurs in the
BLSSM whereas in Sect.~\ref{sect:spectrum} we introduce its particle spectrum. In Sect.~\ref{sect:MassCorr} we
study in particular the Higgs masses. Then, in Sect.~\ref{sect:LHC}, we describe the main manifestations of the
BLSSM at the LHC. We conclude in Sect.~\ref{sect:summary}.

\section{{\large\bf Constructing the BLSSM}\label{sect:construction}}

The particle content of this model
includes the following superfields in addition to those of the Minimal Supersymmetric Standard Model
(MSSM): $(i)$ two SM singlet chiral Higgs superfields $\chi_{1,2}$, whose Vacuum Expectation Values (VEVs)
of their scalar components spontaneously break the
$U(1)_{B-L}$  and $\chi_2$ is required to cancel
the $U(1)_{B-L}$ anomaly; $(ii)$ three sets of SM singlet chiral
superfields, $\nu_i, s_{1_i}, s_{2_i} (i =1,2,3)$, to implement the
IS mechanism (also in order to cancel the $B-L$ anomaly). Tab. \ref{particle-content-BLSSMIS} provides the particle content of the SUSY version of the $B-L$ model with IS (henceforth
 BLSSM-IS for short) as well as the different charge assignments of each Superfield.
\begin{table} [t]
\centering
\begin{tabular}{|c|c|c|c|c|c|}
\hline \hline
Superfield & Spin 0 & Spin \(\frac{1}{2}\) & Generations & $SU(3)_C \times SU(2)_L \times U(1)_Y \times U(1)_{B-L}$ \\
\hline
\(\hat{Q}\) & \(\tilde{Q}\) & \(Q\) & 3
 & \(({{\bf 3},\bf 2},\frac{1}{6},\frac{1}{3}) \) \\
\(\hat{d}^c\) & \(\tilde{d}^c\) & \(d^c\) & 3
 & \(({\bf \overline{3}},{\bf 1},\frac{1}{3},-\frac{1}{3}) \) \\
\(\hat{u}^c\) & \(\tilde{u}^c\) & \(u^c\) & 3
 & \(({\bf 1},{\bf \overline{3}},-\frac{2}{3},-\frac{1}{3}) \) \\
\(\hat{l}\) & \(\tilde{l}\) & \(l\) & 3
 & \(({\bf 1},{\bf 2},-\frac{1}{2},-1) \) \\
\(\hat{e}^c\) & \(\tilde{e}^c\) & \(e^c\) & 3
 & \(({\bf 1},{\bf 1},1,1) \) \\
\(\hat{\nu}^c\) & \(\tilde{\nu}^c\) & \(\nu^c\) & 3
 & \(({\bf 1},{\bf 1},0,1) \) \\
 \(\hat{s}_1\) & \(\tilde{S}_1\) & \(S_1\) & 3
 & \(({\bf 1},{\bf 1},0,2) \) \\
 \(\hat{s}_2\) & \(\tilde{S}_2\) & \(S_2\) & 3
 & \(({\bf 1},{\bf 1},0,-2) \) \\
\(\hat{H}_d\) & \(H_d\) & \(\tilde{H}_d\) & 1
 & \(({\bf 1},{\bf 2},-\frac{1}{2},0) \) \\
\(\hat{H}_u\) & \(H_u\) & \(\tilde{H}_u\) & 1
 & \(({\bf 1},{\bf 2},\frac{1}{2},0) \) \\
\(\hat{\chi}_1\) & \(\chi_1\) & \(\tilde{\chi}_1\) & 1
 & \(({\bf 1},{\bf 1},0,1) \) \\
\(\hat{\chi}_2\) & \(\chi_2\) & \(\tilde{\chi}_2\) & 1
 & \(({\bf 1},{\bf 1},0,-1) \) \\
\hline \hline
\end{tabular}
\caption{Chiral superfields of the BLSSM-IS and their quantum numbers under $SU(3)_C \times SU(2)_L \times U(1)_Y \times U(1)_{B-L}$ .}
\label{particle-content-BLSSMIS}
\end{table}
%
{The superpotential of the leptonic sector in this model is given by
\cite{Elsayed:2011de}
\bea%
W = Y_u\,\hat{u}\,\hat{q}\,\hat{H}_u - Y_d \,\hat{d}\,\hat{q}\,\hat{H}_d\,- Y_e \,\hat{e}\,\hat{l}\,\hat{H}_d \, +
Y_\nu\,\hat{\nu}\,\hat{l}\,\hat{H}_u \, +Y_s\,\hat{\nu}\,\hat{\chi}_1\,\hat{s}_2 +\mu\,\hat{H}_u\,\hat{H}_d \, - {\mu'} \,\hat{\chi}_1\,\hat{\chi}_2\,
\label{superpotential}
\eea}
%
{In order to prevent a possible large mass term $M s_1 s_2$, we assume that the superfields  $\hat{\nu}$, $\chi_{1,2}$  and $s_2$ are even under matter
parity, while $s_1$ is an odd particle. } By assuming a minimal Supergravity (mSugra) inspired universality of parameters at the scale of a Grand Unification Theory (GUT), we obtain that the SUSY soft
breaking Lagrangian is given by
\begin{eqnarray}
-\mathcal{L}_{\tiny\textnormal{soft}} \!\!&\! \!=\! \!&\! \! m_0^2\Big[|\tilde{q}|^2 + |\tilde{u}|^2 + |\tilde{d}|^2 + |\tilde{l}|^2 + |\tilde{e}^{c}|^2 + |\tilde{\nu}^c|^2 + |\tilde{S}_1|^2 + |\tilde{S}_2|^2 + |H_d|^2 + |H_u|^2 +|\chi_1|^2 \n
&+&  |\chi_2|^2\Big]+ \left[ Y_u^A \tilde{q} H_u \tilde{u}^c + Y_d^A \tilde{q} H_d \tilde{d}^c + Y_e^A \tilde{l} H_d \tilde{e}^c + Y_{\nu}^A \tilde{l} H_u \tilde{\nu}^c + Y_s^A \tilde{\nu}^c \chi_1 \tilde{S}_2 \right]\n
&+& \left[ B ( \mu H_1 H_2 + \mu' \chi_1 \chi_2 ) + h.c. \right] + \frac{1}{2} M_{1/2} \left[ \tilde{g}^a \tilde{g}^a + \tilde{W}^a \tilde{W}^a + \tilde{B} \tilde{B} + \tilde{B}' \tilde{B}' + h.c. \right],\nonumber
\end{eqnarray}
where the trilinear terms are defined as $(Y_f^A)_{ij} = ( Y_f A )_{ij}$.

The $B-L$ symmetry is radiatively broken by
the non-vanishing VEVs 
 $\langle{\rm Re} \chi^0_i\rangle=\frac{v'_i}{\sqrt{2}}$ ($i=1,2$)
while the EW one by the non-zero
VEVs $\langle{\rm Re} H_{u,d}^0\rangle=v_{u,d}/\sqrt{2}$, with $v=\sqrt{v^2_u+v^d_2}= 246$ GeV, $v'=\sqrt{v'^2_1+v'^2_2}\simeq {\cal O}(1)$ TeV and the ratio of these VEVs are defined as $\tan \beta = v_u/v_d$ and $\tan \beta' = v'_1/v'_2$ \cite{Khalil:2007dr}. After
$B-L$ and EW symmetry breaking, the neutrino Yukawa
interaction terms lead to the following expression:
\be
{\cal L}_m^{\nu} = m_D\, \bar{\nu}_L \nu^c + M_R\, {\bar {\nu^c}} S_2 +
{\rm {~h.c.}},
\ee
where $m_D=\frac{1}{\sqrt{2}}Y_\nu v_u$ and $ M_R =
\frac{1}{\sqrt 2}Y_{s} v'_1$.
In this framework, the light neutrino masses are related to a
small mass term $\mu_s S^2_2$ in the Lagrangian, with $\mu_s\sim {\cal O}(1)$ {\rm
KeV}, which can be generated at the $B-L$ scale through a
non-renormalisable higher order term $\frac{\chi_1^4 S^2_2}{M^3}$, where $M$  
is the mass of a heavy state whose loop(s) or tree-level tadpole diagrams generate the corresponding higher order term
($M \simeq 10^6$ in our case).Therefore, one finds that the neutrinos mix with the fermionic singlet fields to build up the following $9\times 9$ mass matrix, in the basis
$(\nu_L ,\nu^c, S_2)$:

%
{\be {\cal M}_{\nu}=
\left(%
\begin{array}{ccc}
  0 & m_D & 0\\
  m^T_D & 0 & M_R \\
  0 & M^T_R & \mu_s\\
\end{array}%
\right). %
\label{inverse}
\ee%
The diagonalisation of the mass matrix, Eq. (\ref{inverse}),
leads to the following light and heavy neutrino masses, respectively: %
\begin{eqnarray}%
m_{\nu_l} &=& m_D M_R^{-1} \mu_s (M_R^T)^{-1} m_D^T,\label{mnul}\\
m_{\nu_h}&=& m_{\nu_{H'}} = \sqrt{M_R^2 + m_D^2}. %
\end{eqnarray} }
Thus, one finds that the light neutrino masses can be of order eV, with a TeV scale $M_R$, if $\mu_s \ll
M_R$, and a order one Yukawa coupling $Y_{\nu}$. Such a large coupling is crucial for testing the BLSSM-IS and probing the heavy neutrinos at the LHC. As shown in  \cite{Abdallah:2011ew},
the mixings between light and heavy neutrinos are of order
${\cal O}(0.01)$. Therefore, the decay widths of these heavy neutrinos into SM fermions are sufficiently large. It is worth mentioning that
the second SM singlet fermion, $S_1$, remains light with mass given by%
\be%
m_{S_1} = \mu_s \simeq {\cal O}(1)~ {\rm keV},%
\ee%
where $S_1$ is a sort of inert/sterile neutrino that has no mixing with the active neutrinos. It can therefore be a good
candidate for warm DM as emphasised in Ref. \cite{El-Zant:2013nta}.

\section{{\large\bf Radiative $B-L$ symmetry breaking in the BLSSM-IS}\label{sect:radiative}}

The breaking of $B-L$ can spontaneously occur through the VEV of the scalar field $\chi_{1,2}$ or the right-handed
sneutrino $\tilde{\nu}^c_3$  \cite{Khalil:2007dr}, depending on the initial values of the Yukawa couplings and soft terms involved in the
Renormalisation Group Equations (RGEs)  of the parameters in the scalar potential $V( \chi_1, \chi_2 )$, where
\begin{eqnarray}
V( \chi_1, \chi_2 ) = \mu_1^2 |\chi_1|^2 + \mu_2^2 |\chi_2|^2 - \mu_3^2 ( \chi_1 \chi_2 + h.c. ) + \frac{1}{2} g_{BL}^2 \left( |\chi_2|^2 - |\chi_1|^2 \right)^2,
\end{eqnarray}
where $\mu^2_{1,2} = m^2_{\chi_{1,2}}+ \vert \mu' \vert^2$ , $\mu_3^2=- B' \mu'$ {and $g_{BL}$ is the gauge coupling of $U(1)_{B-L}$.}
The stablitity condition of $V(\chi_1, \chi_2)$ is given by
\begin{equation}
2 \mu_3^2 < \mu_1^2 + \mu_2^2.
\label{4}
\end{equation}

The minimisation of this potential, $\frac{\partial V}{\partial\chi_i}=0,\ i=1,2$, implies that
\begin{eqnarray}
\mu_1^2 = \mu_3^2 \cot \beta' + \frac{M_{_{Z'}}^2}{4} \cos 2 \beta',\label{1}\\
\mu_2^2 = \mu_3^2 \tan \beta' - \frac{M_{_{Z'}}^2}{4} \cos 2 \beta'.\label{2},
\end{eqnarray}
where $M_{Z'}=g_{BL}^2 v'^2$ (no mixing between $U(1)$ and $U(1)_{B-L}$ is assumed here).
From (\ref{1})--(\ref{2}), one gets
\begin{equation}\label{3}
\sin 2\beta'=\frac{2\mu_3^2}{m_{A'_0}^2},
\end{equation}
where $m_{A'_0}^2=\mu_1^2+\mu_2^2$. Note that, again from (\ref{1})--(\ref{2}), we also get
\begin{equation}\label{5}
v'^2=\frac{(\mu_1^2-\mu_2^2)-(\mu_1^2+\mu_2^2)\cos 2\beta'}{2g_{BL}^2\cos 2\beta'}.
\end{equation}
Now we complete our analysis of symmetry breaking. We have
\begin{eqnarray}
V_{11}( v'_1, v'_2 ) & = & 2 \mu_1^2 - 2 g_{BL}^2 ( v'^2_2 - 3v'^2_1 ),\\
V_{12}( v'_1, v'_2 ) & = & - 2 \mu_3^2 - 4 g_{BL}^2 v'_1 v'_2,\\
V_{22}( v'_1, v'_2 ) & = & 2 \mu_2^2 + 2 g_{BL}^2 ( 3v'^2_2 - v'^2_1 ),
\end{eqnarray}
where $V_{ij}=\frac{ \partial^2 V( \chi_1, \chi_2 ) }{ \partial \chi_i \partial \chi_j }$. To show
that the symmetry will be broken spontaneously, we must ensure that the
point $(v'_1,v'_2)=(0,0)$ is not a local minimum of the potential $V$. Since
$\left( V_{11} V_{22} - V_{12}^2 \right)(0,0) = ( 2 \mu_1^2 )( 2 \mu_2^2 )-( 2 \mu_3^2 )^2$
and $V_{11}(0,0)=2\mu_1^2>0$  we should impose a condition to make
$(0,0)$ a saddle point. This condition is
\begin{equation}\label{6}
\mu_1^2\; \mu_2^2 < \mu_3^4.
\end{equation}
It is worth noting that it is impossible to simultaneously fulfil both the
conditions (\ref{4}) and (\ref{6}) for positive values of $\mu_1^2$ and
$\mu_2^2$. However, one should note that the  condition (\ref{6}) is valid at the
$B-L$ symmetry breaking scale, where the running of the RGEs, from the GUT
scale down to the $B-L$ breaking scale, may induce negative squared mass of  $\chi_2$.
At this scale both conditions (\ref{4}) and (\ref{6}) are satisfied and symmetry is spontaneously broken
with stable potential. This can be seen as follows.
%
\begin{figure}[t!]
\begin{center}
\includegraphics[scale=0.85]{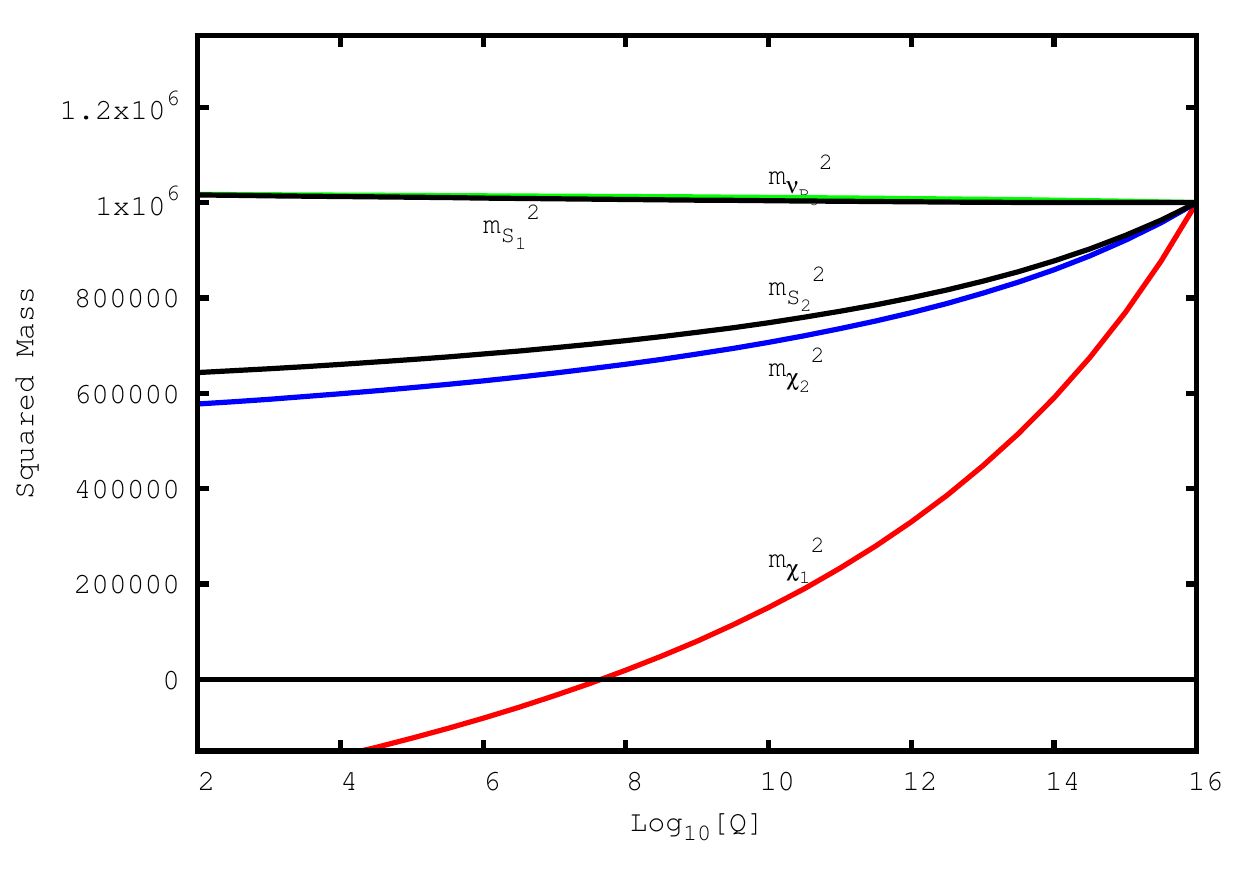}%
\caption{The evolution of the $B-L$ scalar masses from the GUT to the TeV
scale for $m_0 = M_{1/2}= A_0=1$ TeV and $Y_{s_3}\sim {\cal O}(1)$.}
\label{RGE-BLSSM-IS-Fig}
\end{center}
\end{figure}
%
The RGEs for $m^2_{\chi_{1,2}}$ are given by
\begin{eqnarray}
\da{m_{\chi_1}^2} & = & 12 g_{BL}^2 {M}_{BL}^2  - {6Y_{s_3}^2 \left( m_{\chi_1}^2 + m_{\tilde{\nu}^c_3}^2 +  m_{\tilde{S}_{2_3}}^2 + A_{s_3}^2\right)},\\
\da{m_{\chi_2}^2} & = & 12 g_{BL}^2 {M}_{BL}^2,
\end{eqnarray}
{where $t = \ln\left(\frac{M_X^2}{Q^2}\right)$ and $M_{BL}$ is the gaugino mass of $\tilde{B}'$, which is given by $M_{1/2}$ at the GUT scale, $M_X$.} In order to solve these equations, we should take into account all involved RGEs, given in Ref. \cite{Staub:2013tta}.
Fig. \ref{RGE-BLSSM-IS-Fig} reports the result of the running. In plotting this figure, we set the following
mSugra inspired conditions at the
high scale, {e.g.},
$m_0 = M_{1/2}= A_0=100$ GeV and an order one ratio {$Y_{s_3} \simeq M_{R_3}/v'$}. As can be seen from the plot, $m^2_{\chi_1}$ drops rapidly to a negative mass region whereas $m^2_{\chi_2}$ remains positive. Also in Fig. \ref{RGE-BLSSM-IS-Fig}, we plot the scale evolution for the scalar mass {$m_{\tilde{\nu}^c_3}$} as well as for
 {$m_{\tilde{S}_{1_3}}$ and $m_{\tilde{S}_{2_3}}$}. The figure illustrates that they remain positive at the TeV scale. Therefore, the $B-L$ breaking via a non-vanishing VEV for right-handed sneutrinos does not occur in the present framework.

Before closing this section, let us emphasise that in the BLSSM with Type I Seesaw the $B-L$ symmetry is spontaneously broken by the VEV of scalar singlet $\chi_2$ only if the Yukawa coupling $Y_{\nu_R}$ of the term $Y_{\nu_R} \nu_R^c \nu_R^c \chi_2$ is assumed to be degenerate, i.e. $Y_{\nu_R} = Y_0 ~ {\rm diag}\{1,1,1\}$. If one assumes a hierarchical texture, where for instance only $Y_{\nu_{R_3}}$ gives important contributions, one finds that right-handed sneutrino may acquire a VEV before $\chi_2$ and breaks both $B-L$ and $R$-parity. However, here, within the BLSSM-IS, the situation is different. One can show that, even with one non-vanishing Yukawa, $\chi_2$ will acquire a VEV before the right-handed sneutrino such that $B-L$ is spontaneously broken while $R$-parity remain conserved \cite{Khalil:2015}.

\section{{\large\bf The BLSSM-IS Spectrum}\label{sect:spectrum}}
We have seen the evolution of different parameters from the GUT to the $B-L$ scale. Once the $B-L$ symmetry is broken and so is the EW one  too (at the $M_W$ scale), different particles with different quantum numbers can mix and acquire new mass eigenstates. Here, we will focus on the new particles associated with the $B-L$ symmetry, namely, the $Z'$ gauge boson, extra Higgs bosons and right-handed sneutrinos. We shall do so in three separate subsections.

Before proceeding to doing so, we should mention that, in
all the analyses below, we have used the SARAH \cite{Staub:2008uz} and SPheno \cite{Porod:2003um,Porod:2011nf} to build the BLSSM. Furthermore, the matrix-element calculation and event generation were derived from MadGraph 5 \cite{Madgraph5}
and manipulated with MadAnalysis 5 \cite{Madanalysis5}. Finally, notice that all current experimental constraints, from both collider (LEP2, Tevatron and LHC) and flavour  (BaBar,  Belle and LHCb) are taken into account in our numerical
scans. Also DM and neutrino mass and coupling constraints are enforced throughout.

\subsection{{\large\bf The $Z'$ Gauge Boson in the BLSSM-IS}}
The $U(1)_Y$ and $U(1)_{B-L}$ gauge kinetic mixing can be absorbed in the covariant derivative redefinition, where the gauge coupling matrix will be transformed as follows:
\be
G = \left(\begin{array}{cc}
  g_{_{YY}} & g_{_{YB}}\\
  g_{_{BY}} & g_{_{BB}}\\
\end{array}%
\right) ~~ \Longrightarrow ~~ \tilde{G} = \left(\begin{array}{cc}
  g_1& \tilde{g}\\
  0 & g_{_{BL}}\\
\end{array}%
\right) , %
\ee
where
{\bea
g_1 &=& \frac{g_{_{YY}} g_{_{BB}} - g_{_{YB}} g_{_{BY}}}{\sqrt{g_{_{BB}}^2 + g_{_{BL}}^2}},\\
g_{_{BL}} &=&\sqrt{g_{_{BB}}^2 + g_{_{BY}}^2},\\
\tilde{g} &=& \frac{g_{_{YB}} g_{_{BB}} + g_{_{BY}} g_{_{YY}}}{\sqrt{g_{_{BB}}^2 + g_{_{BY}}^2}}.
\eea}
 In this basis, one finds
\be
M_Z^2 = \frac{1}{4} (g_1^2 +g_2^2) v^2,  ~~~~ M_{Z'}^2 = g_{_{BL}}^2 v'^2 + \frac{1}{4} \tilde{g}^2 v^2 .
\ee
Furthermore, the mixing angle between $Z$ and $Z'$ is given by
\be
\tan 2 \theta' = \frac{2 \tilde{g}\sqrt{g_1^2+g_2^2}}{\tilde{g}^2 + 16 (\frac{v'}{v})^2 g_{BL}^2 -g_2^2 -g_1^2},
\ee

\subsection{{\large\bf The Higgs Bosons in the BLSSM-IS}}

The gauge kinetic term induces mixing at tree level between the $H^{{0}}_{1,2}$ and $\chi^{{0}}_{1,2}$
states in the BLSSM scalar potential. Therefore, the minimisation conditions of this potential at tree level lead to the following relations \cite{florian2012}:
\bea
B \mu &=& -\frac{1}{8} \Big[-2 \tilde{g}g_{BL} v'^2 \cos2\beta' + 4 m_{H_1}^2 - 4 m_{H_2}^2\nonumber\\
      &+& (g_1^2 + \tilde{g}^2 + g_2^2) v^2 \cos 2 \beta \Big] \tan 2 \beta ,\label{Bmu}\\
B {\mu'} &=& \frac{1}{4} \Big[-2 g^2_{BL} v'^2 \cos2\beta' + 2 m_{\chi_1}^2 - 2 m_{\chi_2}^2\nonumber \\
      &+&\tilde{g}g_{BL} v^2 \cos 2 \beta \Big] \tan 2 \beta',
\eea
{where $\tan{\beta}=\frac{v_2}{v_1}$ and $\tan{\beta'}=\frac{v'_1}{v'_2}$.} Note that, with non-vanishing $\tilde{g}$, the $B\mu$ parameter depends on $v'$ and  the sign of $\cos 2\beta'$. We may have both constructive and destructive interference between the first term and other terms in Eq.~(\ref{Bmu}). In general, we find that the typical value of $B \mu$ is  of order TeV.

To obtain the masses of the physical neutral Higgs bosons, one makes the usual redefinition of the Higgs fields, i.e.,
$H_{1,2}^0 = {\frac{1}{\sqrt{2}}}(v_{1,2} + \sigma_{1,2} + i \phi_{1,2}) $ and
$\chi_{1,2}^0 ={\frac{1}{\sqrt{2}}}(v'_{1,2} + \sigma'_{1,2}  + i \phi'_{1,2})$.
The real parts correspond to the CP-even Higgs bosons and the imaginary parts correspond to the CP-odd Higgs bosons. The squared-mass matrix of the BLSSM CP-odd neutral Higgs fields at tree level, in the basis $(\phi_1,\phi_2,\phi'_1,\phi'_2)$,  is given by\\
\begin{equation}
m^2_{A,A'} =
\left(\begin{array}{cccc}
B \mu \tan\beta & B \mu  & 0 & 0 \\
B \mu  & B \mu  \cot\beta & 0 & 0 \\
 0 & 0  & B {\mu'}  \tan\beta' & B {\mu'}  \\
 0 & 0  & B {\mu'} & B {\mu'} \cot\beta'
\end{array}
\right) \,.
\label{eq:mA2}
\end{equation}
It is clear that the MSSM-like CP-odd Higgs $A$ is decoupled from the BLSSM-like one $A'$ (at tree level). However, due to the dependence of $B_\mu$ on $v'$, one may find $m_A^{{2}} = \frac{2B_\mu}{\sin 2 \beta} \sim m_{A'}^{{2}} =\frac{2 B {\mu'}}{\sin2 \beta'} \sim {\cal O}(1$ TeV).

The squared-mass matrix of the BLSSM CP-even neutral Higgs fields at tree level, in the basis $(\sigma_1,\sigma_2,\sigma'_1,\sigma'_2)$,  is given by
\begin{equation}
 M^2 = \left( \begin{array}{cc}
 			     M^2_{h{H}} ~ &~  M^2_{hh'} \\  \\
                              M^{2^{{T}}}_{hh'} ~ &~  M^2_{h'{H}'} \end{array} \right),
\label{BLMH}
\end{equation}
where $M^2_{h{H}} $ is the usual MSSM neutral CP-even Higgs mass matrix, which leads to the SM-like Higgs boson with mass, at one loop level, of order 125 GeV and a heavy Higgs boson with mass $m_H \sim m_A \sim {\cal O}(1$ TeV).
In this case, the BLSSM matrix $M^2_{h'{H}'}$  is given by\\
\be
M^2_{h'{H}'}=
\left( \begin{array}{cc}
               m^2_{A'} c^2_{\beta'} + g^2_{BL} v'^2_1 &-\frac{1}{2} m^2_{A'} s_{2\beta'} - g^2_{BL} v'_1 v'_2 \\
                    \\
              -\frac{1}{2} m^2_{A'}s_{2\beta'} - g^2_{BL} v'_1 v'_2 & m^2_{A'} s^2_{\beta'} + g^2_{BL} v'^2_2
              \end{array}\right),
\ee
where $c_x=\cos(x)$ and $s_x=\sin(x)$. Therefore, the eigenvalues of this mass matrix are given by
\bea
{m}^2_{h',H'} = \frac{1}{2} \Big[ ( m^2_{A'} + M_{{Z'}}^2 )
 \mp\sqrt{ ( m^2_{A'} + M_{{Z'}}^2 )^2 - 4 m^2_{A'} M_{{Z'}}^2 \cos^2 2\beta' }\;\Big].
\eea
If $\cos^2{{2}\beta'} \ll 1$, one finds that the lightest $B-L$ neutral Higgs state is given by %
\be%
{m}_{h'}\; {\simeq}\; \left(\frac{m^2_{A'} M_{{Z'}}^2 \cos^2 2\beta'}{{m^2_{A'}+M_{{Z'}}^2}}\right)^{\frac{1}{2}} \simeq {\cal O}(100~ {\rm GeV}).%
\ee%
The mixing matrix $M_{hh'}^2$ is proportional to $\tilde{g}$ and can be written as \cite{florian2012}
\be
M^2_{hh'}=  \frac{1}{2}\tilde{g} g_{BL} \left( \begin{array}{cc}
              v_1 v'_1 & -  v_1 v'_2\\
                    \\
               ~ - v_2 v'_1 &  ~  v_2 v'_2
              \end{array}\right).
\ee
For a gauge coupling $g_{BL} \sim \vert \tilde{g}\vert  \sim {\cal O}(0.5)$, these off-diagonal terms are about one order of magnitude smaller than the diagonal ones. However, they are still crucial for generating interaction vertices between the genuine BLSSM Higgs bosons and the MSSM-like Higgs states. Note that the mixing gauge coupling constant, $\tilde{g}$, is a free parameter that can be positive or negative \cite{florian2012}.

\begin{figure}[t]
\begin{center}
\epsfig{file=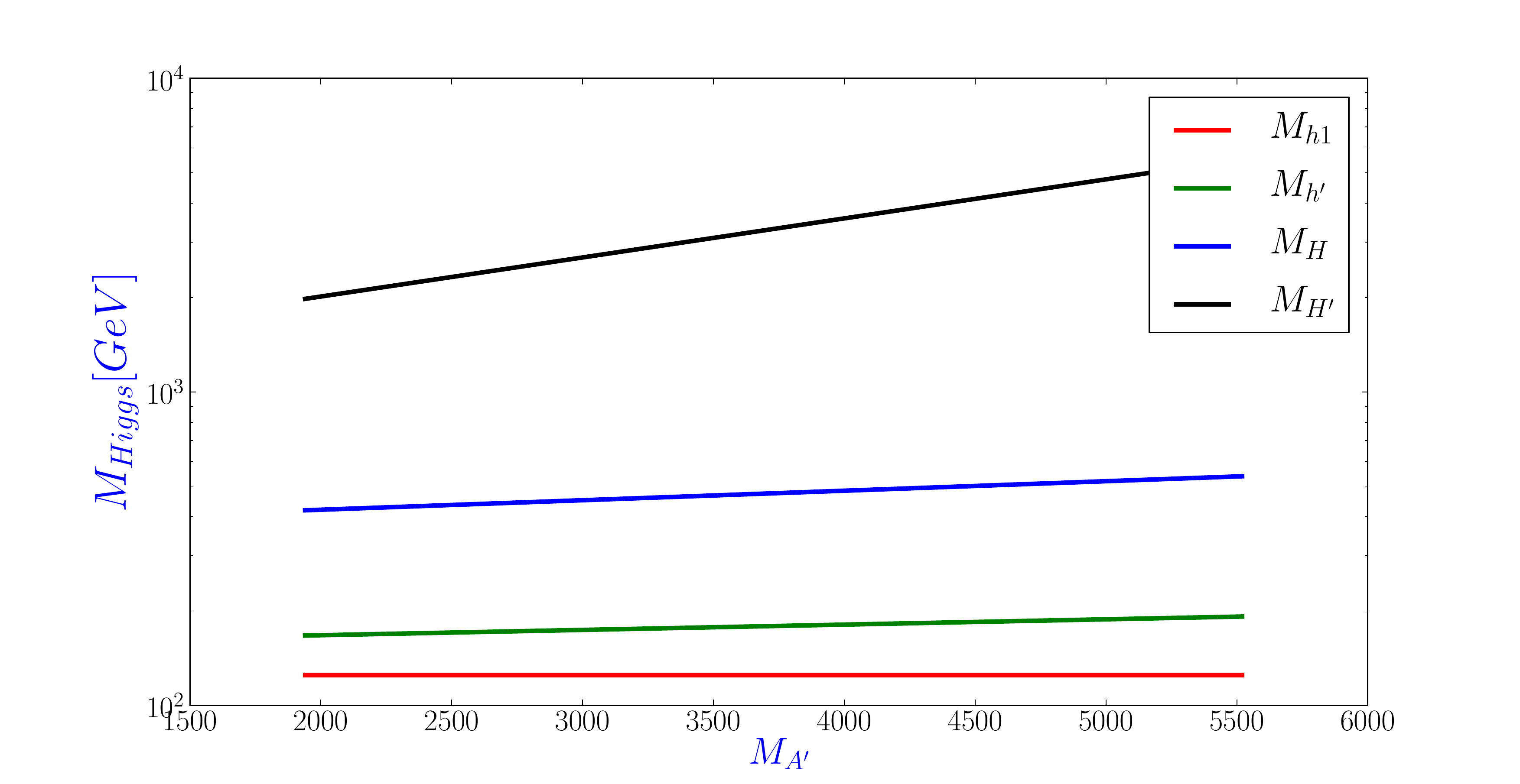,height=7cm,width=11cm,angle=0}
\caption {The BLSSM-IS CP-even Higgs masses versus  $m_{A'}$ for $g_{BL}=0.4$ and $\tilde{g}=-0.4$.}
\label{fig11}
\end{center}
\end{figure}

In Fig.~\ref{fig11}, we show the masses of the four CP-even Higgs bosons in the BLSSM for $g_{BL}=0.4$ and $\tilde{g}=-0.4$ .  In this plot we fix the lightest MSSM Higgs boson mass to be of order 125 GeV. As can be seen from this figure, as intimated, one of the BLSSM Higgs bosons, $h'$, can be the second lightest Higgs boson ($\sim 137$ GeV).  Both $H$ and $H'$ are instead quite heavy (since both $m_A$ and $m_{A'}$ are of order TeV).

This sets the stage for the hypothesis made in Ref.~\cite{Abdallah:2014fra} (see also \cite{Hammad:2015eca}),
wherein, motivated by a $\sim 2.9 \sigma$ excess recorded by the ATLAS and CMS experiments at the LHC around a mass of order $\sim 137$ GeV in $ZZ \to 4l$ (CMS) and $\gamma \gamma$ (both) samples
\cite{ATLAS-CONF-2013-012,Aad:2014eha,Chatrchyan:2013mxa,CMS:2013wda}, is was shown that a double Higgs peak structure can be generated in the BLSSM, with CP-even Higgs boson masses at $\sim 125$ and $\sim 137$ GeV, a possibility instead precluded to the
MSSM.

Before proceeding in this respect, though, two remarks are in order: firstly, if $\tilde{g}=0$, the coupling of the BLSSM lightest Higgs state, $h'$, with the SM particles will be significantly suppressed ($\leq 10^{-5}$ relative to the SM strength), so that, in order to account for possible $h'$ signals at the LHC, this parameter ought to be sizable; secondly, in both cases of vanishing and non-vanishing $\tilde{g}$, one may fine-tune the parameters and get a light $m_A$, which leads to a MSSM-like CP-even Higgs state, $H$, with $m_H \sim 137$ GeV. However, it is well known that in the MSSM the coupling $HZZ$ is suppressed with respect to the corresponding one of the SM-like Higgs particle by one order of magnitude due to the smallness of $\cos(\beta -\alpha)$,
where $\sin(\beta -\alpha) \sim 1$. In addition, the total decay width of $H$ is larger than the total decay width of the SM-like Higgs,
$h$, by at least one order of magnitude, because it is proportional to $(\cos \alpha/\cos\beta)^2$, which is essentially the square of the coupling of $H$ to the bottom quark. Therefore, the MSSM-like heavy Higgs signal $(pp \to  H \to  ZZ \to 4l)$ has a very suppressed cross section and thus cannot be a candidate for light Higgs signals at the LHC.

In the light of this, we will focus in the next section on the lightest BLSSM CP-even Higgs, $h'$, as a possible candidate for the second Higgs peak seen by CMS in both $ZZ\to 4l$ and $\gamma\gamma$, see  Refs.~\cite{Chatrchyan:2013mxa,CMS:2013wda}. However, before doing so, we ought to setup
appropriately the BLSSM parameter space, in order to find such a solution.
As mentioned above, the recent results from CMS  indicate a $\sim 2.9 \sigma$ hint of a second Higgs boson at $137$  GeV. Herein, for definiteness, we consider $m_{h'} = 136.5$ GeV as reference BLSSM point.

\begin{figure}[t]
\begin{center}
\epsfig{file=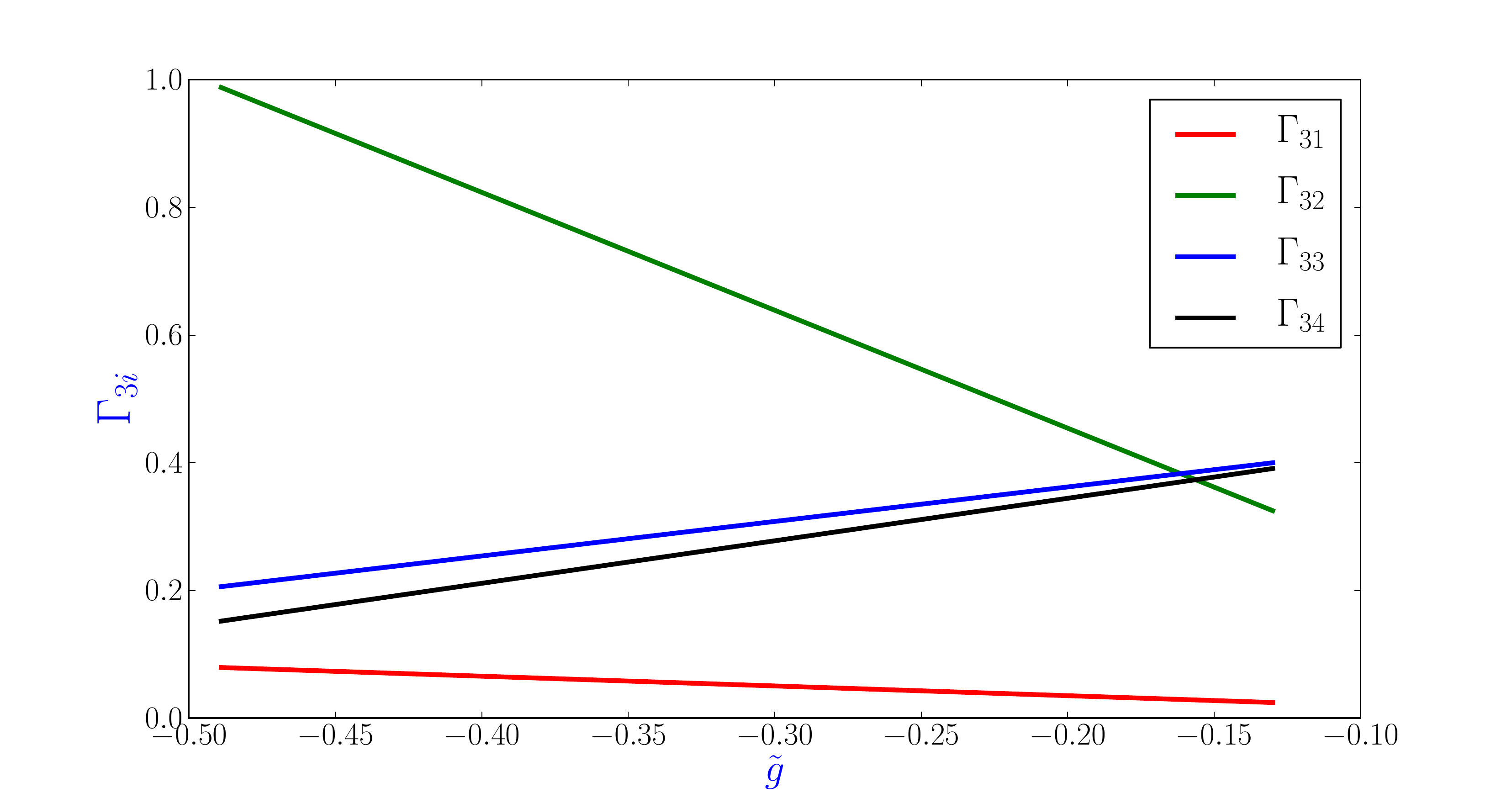,height=7cm,width=11cm,angle=0}
\caption {The mixing of $h'$, $\Gamma_{3i}$, versus the gauge kinetic mixing, $\tilde{g}$.}
\label{fig5}
\end{center}
\end{figure}

As emphasised above, in the BLSSM, it is quite natural to have two light CP-even Higgs bosons, $h$ and $h'$, with mass $125$ GeV and $\sim 137$ GeV, respectively. The CP-even neutral Higgs mass matrix in Eq.~(\ref{BLMH}) can be diagonalised by a unitary transformation:
\be
{\Gamma}~ M^2 ~ \Gamma^\dag = {\rm diag}\{m_h^2, m_H^2, m_{h'}^2, m_{H'}^2\}.
\ee
The mixing elements $\Gamma_{32}$ and $\Gamma_{31}$ are proportional to $\tilde{g}$ and they identically vanish if $\tilde{g}=0$, as one can see in Fig.~\ref{fig5}. Also, in this limit, $\Gamma_{11}$ and $\Gamma_{12}$ approach $\sin \alpha$ and $\cos\alpha$, respectively, where $\alpha$ is the usual CP-even Higgs mixing angle in the MSSM.

The lightest eigenstate $h$ is the SM-like Higgs boson, for which we will fix its mass to be exactly 125 GeV. As mentioned, numerical scans of the BLSSM parameter space  confirm that the $h'$ state can then be the second light Higgs boson with mass of ${\cal O}$(137 GeV). The other two CP-even states, $H$ and $ H'$, are heavy (of ${\cal O}(1)$ TeV). The $h'$ can be written in terms of  gauge eigenstates as
\be
h' = \Gamma_{31} ~\sigma_1 + \Gamma_{32} ~\sigma_2 + \Gamma_{33}~ \sigma'_1  +\Gamma_{34}~ \sigma'_2.
\ee
Thus, the couplings of the $h'$ with up- and down-quarks  are given in terms of the
elements of the  $\Gamma$ mixing matrix
by
\be%
h'\, u\, \bar{u} :~ {-i}~\frac{m_u}{v} \frac{\Gamma_{{32}}}{\sin \beta}, ~~~~~  h'\, d\, \bar{d} : ~ {-i}~\frac{m_d}{v} \frac{\Gamma_{{31}}}{\cos \beta}.
\ee
Similarly, one can derive the $h'$ couplings with the ${W^{+}}{W^-}$ and $Z{Z}$ gauge boson pairs:
\bea%
h'\,W^+\,W^- &:& {i}~g_{{2}} M_W \left(\Gamma_{{32}} \sin \beta + \Gamma_{{31}} \cos\beta\right),\n
h' Z Z &:&\frac{i}{2}\Big[4 g_{BL}\sin^2{\theta'}\left(v'_1\Gamma_{32} + v'_2\Gamma_{31}\right)\n
&+&\left(v_2\Gamma_{32} + v_1\Gamma_{31}\right)\left(g_Z \cos{\theta'}-\tilde{g}\sin{\theta'}\right)^2\Big].\nonumber
\eea
{Since $\sin{\theta'}\ll 1$, the coupling of the $h'$ with $ZZ$, $g_{h'ZZ}$, will be as follows:}
\be
g_{h'ZZ}\simeq i~g_Z\, M_Z  \left(\Gamma_{32} \sin{\beta}+ \Gamma_{31} \cos{\beta}\right),
\ee
where $g_Z=\sqrt{g_1^2+g_2^2}$.

\subsection{{\large\bf Sneutrino Masses in the  BLSSM-IS}}

\begin{figure}[t!]
\begin{center}
\includegraphics[height=5.85cm, width=8.95cm]{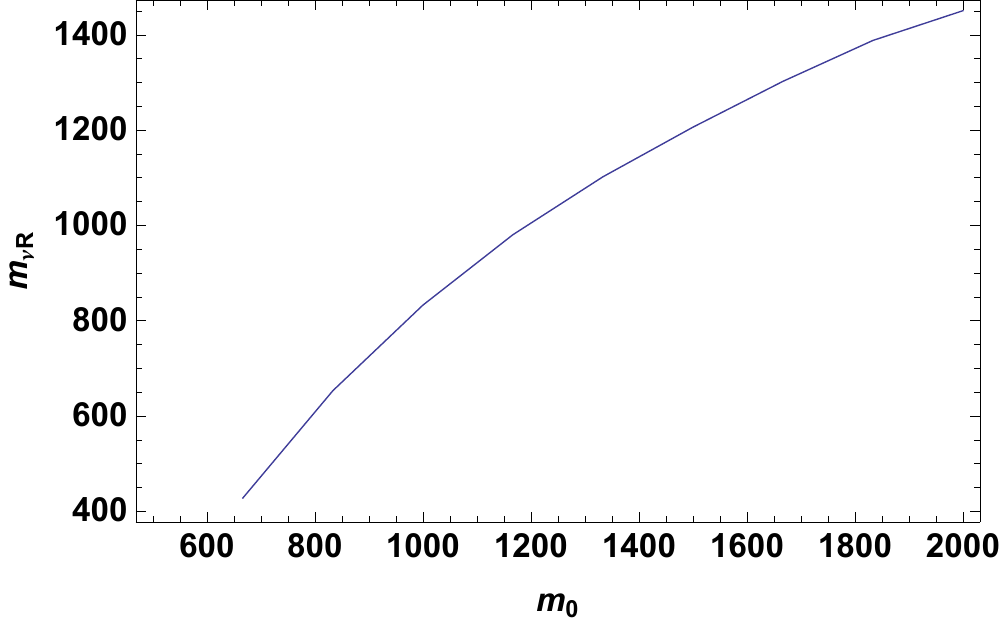}%
\caption{The lightest sneutrino mass as function of $m_0$ for $m_{1/2}=1.5$ TeV and $A_0 = 2.5$ TeV, so that the SM-like Higgs boson mass is within experimental limits. }
\label{snuR-Fig}
\end{center}
\end{figure}

Now we turn to the sneutrino mass matrix.  If we write $\tilde{\nu}_{L,R}$ and  $\tilde{S}_2$ as $\tilde{\nu}_{L,R} = \frac{1}{\sqrt{2}}(\phi_{L,R} + i \sigma_{L,R})$ and
$\tilde{S}_{2} = \frac{1}{\sqrt{2}}(\phi_{S} + i \sigma_{S})$, then we get the following mass matrix for the CP-odd sneutrinos:
\begin{equation}
m^2_{\tilde{\nu}^i} = \left(
\begin{array}{ccc}
m_{\sigma_L \sigma_L} &m^T_{\sigma_L\sigma_R} &\frac{1}{2} v_2 v'_1 Re(Y_{\nu}^{T}  Y_s^*) \\
m_{\sigma_L\sigma_R} &m_{\sigma_R\sigma_R} &m^T_{\sigma_R\sigma_S}\\
\frac{1}{2} v_2 v'_1 Re(Y_{s}^{T}  Y_\nu^*)  &m_{\sigma_R\sigma_S} &m_{\sigma_S \sigma_S}\end{array}
\right),
 \end{equation}
 where $m_{\sigma_L \sigma_L}$, $m_{\sigma_L\sigma_R}$, $m_{\sigma_R\sigma_S} $, and $m_{\sigma_S\sigma_S}$ are given in Ref. \cite{Elsayed:2011de} and  are proportional
 to $v^2$, $v A_0$, $v'^2$, $v' \mu'$, and $v'^2$, respectively. The mass matrix for the CP-even sneutrino ($m_{\tilde{\nu}^{R}}$) is obtained by changing $\sigma_{L,R,S} \to \phi_{L,R,S}$. The sneutrino mass eigenstates can be obtained by diagonalising the mass matrices as

\begin{equation}
 U_{\tilde{\nu}^{i}}m_{\tilde{\nu}^{i}}^{2}U_{\tilde{\nu}^{i}}^{\dagger}=m^{dia}_{\tilde{\nu}^{i}}~,\hspace{0.3cm}
 U_{\tilde{\nu}^{R}}m_{\tilde{\nu}^{R}}^{2}U_{\tilde{\nu}^{R}}^{\dagger}=m^{dia}_{\tilde{\nu}^{R}}.
 \label{nudiag}
 \end{equation}
\noindent
However, the diagonalisation of these matrices is not an easy task and can only be performed numerically. It turns out that the mass of the lightest CP-odd sneutrino, $\tilde{\nu}_I$, is almost equal to the mass of the lightest CP-even sneutrino, $\tilde{\nu}_R$. Both $\tilde{\nu}_{I,R}$ are generated from the mixing between $\tilde{\nu}_R$ and $\tilde{S}_2$. The mass of these lightest sneutrinos can be {\textcolor{blue}{in the range 400--1400}} GeV, as shown in Fig. \ref{snuR-Fig}.

\section{{\large\bf (S)Neutrino Corrections to the Lightest Higgs Boson Mass}\label{sect:MassCorr}}

In this section, we calculate the one-loop radiative corrections due to right-handed (s)neu\-trinos to the mass of the lightest Higgs boson when the latter is SM-like. We show that such effects is in the range $400 -1400$ GeV, thereby giving an
absolute upper limit on such a mass around 170 GeV \cite{Elsayed:2011de}. The importance of this result from a phenomenological point of view resides in the fact that this enhancement greatly reconciles theory and experiment, by alleviating the so-called `little hierarchy problem' of the minimal SUSY realisation, whereby the currently measured mass of the SM-like Higgs mass is very near its absolute upper limit predicted theoretically, of 130 GeV.

It is important to note that, unlike the squark sector, where only the third generation (stops) has a large Yukawa coupling with the Higgs boson, hence giving the relevant correction to the Higgs mass, all three generations of the (s)neutrino sector may lead to important effects since the neutrino Yukawa couplings are generally not hierarchical. Also, due to the large mixing between the right-handed neutrinos $N_i$ and $S_{2_j}$ \cite{Khalil:2011tb}, all the right-handed sneutrinos $\tilde{\nu}_H$ are coupled to the Higgs boson $H_2$, hence they can give significant contribution to the Higgs mass correction. In this respect, it is useful to note that the stop effect is due to the running of 12 degrees of freedom (3 colors times 2 charges times 2 for left and right stops) in the Higgs mass loop corrections, just like in the case
of right-handed sneutrinos for which there are also 12 degrees of freedom (3 generations times 4 eigenvalues).

To calculate the (s)neutrino corrections to the lightest Higgs mass, we computed, in a previous section, the explicit form of the sneutrino masses, while for the neutrino mass expressions, which are well known, we refer the reader to \cite{Elsayed:2011de}.
Due to one generation of neutrinos and sneutrinos, the one-loop radiative correction to the effective potential is given by the relation
\begin{eqnarray}
\Delta V_{\nu,\nn}=\frac{1}{64\pi^2}\Big[\sum_{i=1}^6 m_{\nn_i}^4\Big(\log\frac{m^2_{\nn_i}}{Q^2}-\frac{3}{2}\Big) - 2
\sum_{i=1}^3 m_{\nu_i}^4\Big(\log\frac{m^2_{\nu_i}}{Q^2}-\frac{3}{2}\Big)\Big].
\end{eqnarray}
The first sum runs over the sneutrino mass eigenvalues, while the second sum runs over the neutrino masses (with vanishing $m_{\nu_1}$). In case of degenerate diagonal Yukawa couplings, one finds that the total $\Delta V_{\nu,\nn}$ is given by three times the value of $\Delta V_{\nu,\nn}$ for one generation. This factor then compensates the colour factor of (s)top contributions.

Therefore, the genuine $B-L$ correction to the CP-even Higgs mass matrix, due to the (s)neutrinos, at the scale $\hat{Q}$ at which $\frac{\partial (\Delta V_{\nu,\nn})}{\partial v_k}=0$, is given by
\begin{eqnarray}
\Delta M_{ij}^2 = \frac{1}{2}\;\frac{\partial^2 (\Delta V_{\nu,\nn})}{\partial v_i \partial v_j}.
\end{eqnarray}
It follows that (see \cite{Elsayed:2011de} for details)
{\fontsize{10}{10}\selectfont
\begin{eqnarray}
\frac{\partial^2 (\Delta V_{\nu,\nn})}{\partial v_k \partial v_\ell} & = & \frac{1}{32\pi^2}\sum_i(-1)^{2J_i}(2J_i+1)\;\frac{\partial m_i^2}{\partial v_k}\,\frac{\partial m_i^2}{\partial v_{\ell}}\log\frac{m_i^2}{\hat{Q}^2}\n
& = & \frac{1}{32\pi^2}\left[4 (2Y_\nu^2 v_2)(2Y_\nu^2 v_2)\delta_{k,2}\delta_{\ell,2}\log\frac{m_{\nn_i}^2}{Q_0^2}- 2 \left(2 (2Y_\nu^2 v_2)(2Y_\nu^2 v_2) \delta_{k,2}\delta_{\ell,2}\log\frac{m_{\nu_i}^2}{Q_0^2} \right)\right]\n
& = & \frac{m_D^4}{2\pi^2 v_2^2}\log\left(\frac{m_{\nn_i}^2}{m_{\nu_i}^2}\right) \delta_{k,2}\delta_{\ell,2}.
\end{eqnarray}}
That is, we have
\begin{eqnarray}
\Delta M_{11}^2 & = & \Delta M_{12}^2\ =\ \Delta M_{21}^2\ =\ 0,\\
\Delta M_{22}^2 & = & \frac{m_D^4}{4\pi^2 v_2^2}\,\log\frac{m_{\nn_i}^2}{m_{\nu_i}^2}.
\label{deltanu}
\end{eqnarray}
Therefore, the complete one-loop squared-mass matrix of CP-even Higgs bosons will be given by $M_{\small\rm{tree}}^2+~\Delta M^2$, with
\begin{eqnarray}
\Delta M^2 =\left(%
\begin{array}{cc}
  0 & 0\\
  0 & \delta_{t}^2 + \delta_{\nu}^2\\
  \end{array}%
\right),%
\end{eqnarray}
where $\delta_t^2$ refers to the (s)top contribution presented in Eq.~(1) of  \cite{Elsayed:2011de}
and $\delta_\nu^2$ is the (s)neutrino correction given in Eq. (\ref{deltanu}). In this case, the
lightest Higgs bosons mass is given by
\begin{eqnarray}
m_h^2 = \frac{M_A^2 + M_Z^2 + \delta_t^2 + \delta_\nu^2}{2} \left[1 - \sqrt{1 - 4\frac{M_Z^2
M_A^2 \cos^2 2\beta + (\delta_t^2 + \delta_\nu^2) (M_A^2 \sin^2 \beta + M_Z^2 \cos^2 \beta)}{(M_A^2 + M_Z^2 + \delta_t^2 + \delta_\nu^2)^2}}\right].
\end{eqnarray}

For $M_A\gg M_Z$ and $\cos2\beta\simeq 1$, one finds that
\begin{equation}
m_h^2 \simeq M_Z^2+\delta_t^2+\delta_\nu^2.
\end{equation}
If $\tilde{m}\simeq{\cal O}(1)$ TeV, $Y_\nu\simeq{\cal O}(1)$ and $M_N\simeq{\cal O}(500)$ GeV, one gets that $\delta_\nu^2\simeq{\cal O}(100~{\rm GeV})^2$, thus the Higgs mass is of order $\sqrt{(90)^2+{\cal O}(100)^2+{\cal O}(100)^2}$ GeV $\simeq 170$ GeV.
%

\begin{figure}[t]
\begin{center}
\includegraphics[scale=1.0]{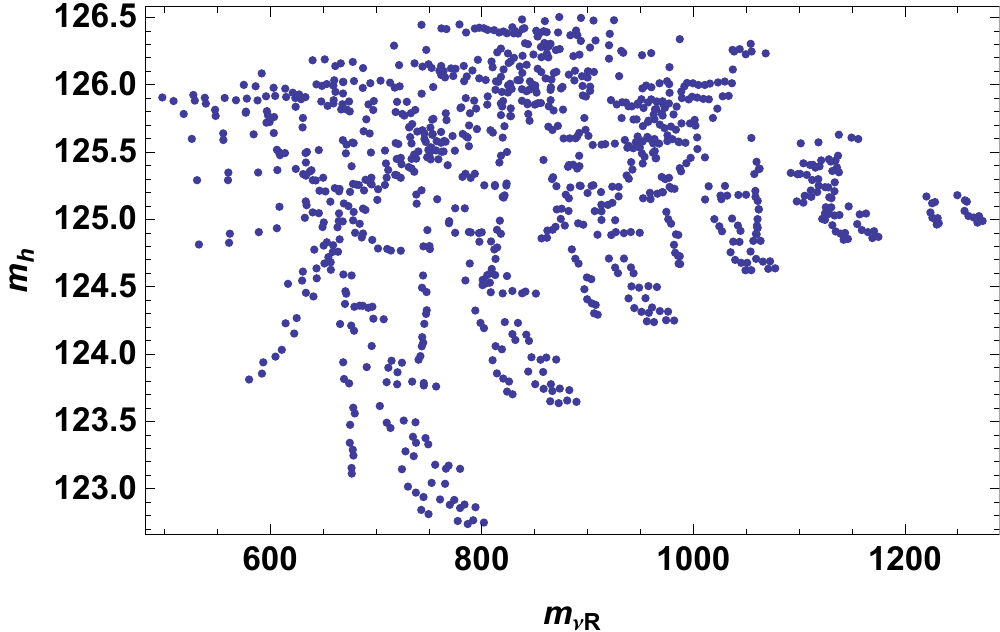}
\vspace{-0.5cm}\caption{Lightest
Higgs boson mass versus the lightest sneutrino mass.} %
\end{center}
\label{mh-mnu}
\end{figure}
%

In Fig.~\ref{mh-mnu} we present the Higgs mass, $m_h$, as a function of the sneutrino mass, $m_{\tilde{\nu}}$, for $m_0\in[500,1000]$ GeV and $Y_{\nu_i}$ couplings varying from 0.1 to 0.4. As can be seen, in our model the Higgs mass can be easily within the experimental limits.  In particular, we have employed the Higgs mass bound as $123~{\rm GeV}\leq m_{h} \leq 127~{\rm GeV}$ \cite{Aad:2012tfa,Chatrchyan:2012xdj}, where we take into account about 2 GeV uncertainty in the Higgs boson mass due to the theoretical uncertainties in the calculation of the minimum of the scalar potential,  and the experimental uncertainties in $m_{t}$ and $\alpha_{s}$. Finally, notice that all the data points collected satisfy the requirement of radiative EWSB.


\section{{\large\bf LHC Signatures of the BLSSM-IS}\label{sect:LHC}}

\subsection{\large\bf Search for the BLSSM-IS $Z'$}

\begin{figure}[t]
\begin{center}
\epsfig{file=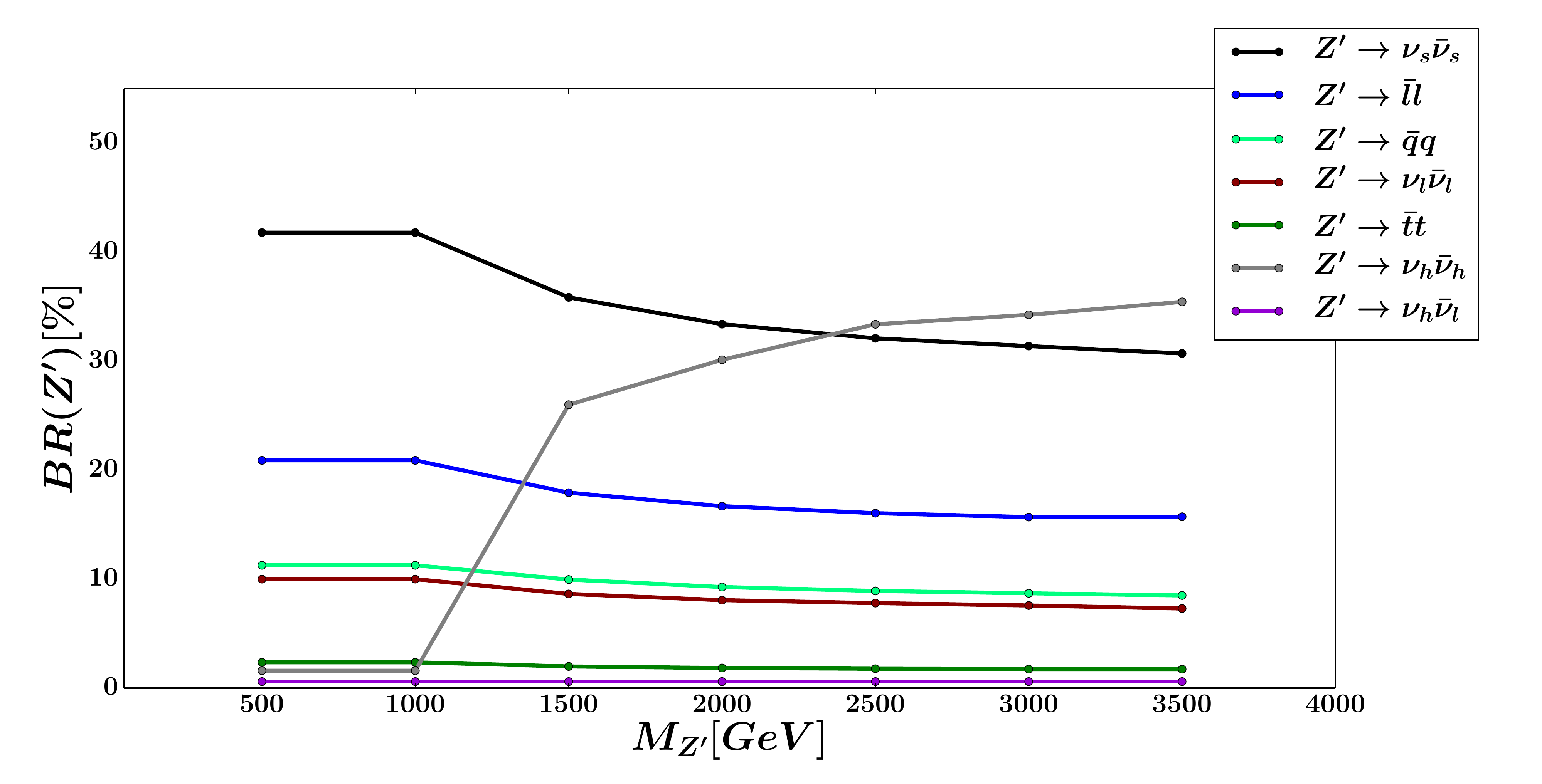,height=7cm,width=11cm,angle=0}
\caption {BRs of the $Z^\prime$ decays in the BLSSM-IS as a function of $M_{Z'}$
(note that  fermion species are  summed over), for $g_{BL} =0.5$ and $\tilde{g}=0.1$.}
\label{fig1}
\end{center}
\end{figure}

Now we study the signatures of the extra neutral gauge boson $Z^\prime$ in the BLSSM-IS at the CERN machine\footnote{In fact, we assume here that all SUSY particles (including sneutrinos) are large enough so that the $Z'$ cannot decay into
these, thereby implying that our analysis can be applied also to the standard $B-L$ scenario. We will look at some SUSY effects later on in this review.}.  The possibility of a $Z^\prime$ decay into a pair of heavy (inert) neutrinos would increase the total decay width of
the $Z'$. Therefore, the Branching Ratio (BR) of $Z' \to l^+ l^-$  ($l=e,\mu$), the prime $Z'$ signal at the LHC,
 is suppressed with respect to the prediction of, {e.g.}, the Sequential Standard Model (SSM), which is usually considered as benchmark in experimental searches for a $Z'$. Fig. \ref{fig1} shows the BRs of all $Z^\prime$ decays. Note that we have assumed that the sfermion are quite heavy so that the $Z'$ decay is dominated by SM particles and light inert/sterile neutrinos.
According to this plot, the BRs of the non-SUSY $Z^\prime$ decays are given by \cite{Abdelalim:2014cxa}%
\bea
&&\sum_{l} BR (Z^\prime \rightarrow l \bar{l}) \sim  16.1\%, \nonumber\\
&&\sum_{\nu_l} BR (Z^\prime \rightarrow \nu_l \bar{\nu_l}) \sim  7.8\%, \nonumber\\
&&\sum_{q} BR(Z^\prime \rightarrow q \bar{q}) \sim 8.92\%, \nonumber\\
&&\sum_{\nu_h} BR (Z^\prime \rightarrow \nu_h \bar{\nu}_h) \sim 33.4\%, \nonumber\\
&&\sum_{\nu_s} BR (Z^\prime \rightarrow \nu_s \bar{\nu}_s) \sim 32.1 \% ,\nonumber\\
&&\sum_{\nu_l,\nu_h} BR (Z^\prime \rightarrow \nu_l \bar{\nu}_h) \sim 0.6 \%,
\eea
where $l$, $q$, and $\nu_h$  refer to  the charged leptons, the six quarks, and  the six heavy neutrinos respectively.
Whereas $\nu_s$ stands for the three inert neutrinos. In this example we have
assumed $M_{Z'}=2.5$ TeV, $g_{BL} =0.5$, $\tilde{g}=0.1$ and heavy neutrino masses are set at $200$, $430$ and $600$ GeV, respectively.

It is worth noting that, in our model, the $Z^\prime$ cross sections ($\sigma$'s) that were used to derive the ATLAS and CMS current mass limit could be simply rescaled by a factor of $(g_{B-L}/g_{Z})^2 \times (1 - {\rm BR}(Z^\prime\to{\rm new\, decay\,
channels}))$. If $g_{B-L}=g_Z$ and BR$(Z^\prime\to{\rm new\, decay\,
channels})=0$, this reproduces the SSM cross sections that were used by ATLAS and CMS. Considering the scaling of cross sections, the current $Z^\prime$ mass limits will be lowered by a factor of $\sigma_{B-L}(Z^\prime \to ll)/\sigma_{\rm SSM}(Z^\prime \to ll)$. This result is consistent with the conclusion of Ref. \cite{Arcadi:2013qia}.

 If $M_Z^\prime = 1000$ GeV were considered,  BR$(Z' \to  l^+ l^-) \sim 14\%$ could be achieved (e.g., through onsetting $Z'$ decays into inert/sterile neutrinos),
in which case $\sigma \times {\rm BR} = 16  $ fb when $g_{B-L} = g_Z=0.188$ and $\sigma\times {\rm BR} = 82  $ fb when $g_{B-L} = 0.5$, while in the SSM the BR$(Z' \to  l^+ l^-) \sim 7.6\%$ giving $\sigma\times{\rm BR}$ $= 340$ fb for both electron and muon channels. In this respect, the experimental limit $M_{Z'}  \gsim 2.5$ TeV by ATLAS \cite{AtlasResult} ($2.8$ TeV \cite{cmsResult} by CMS) will be lowered, because of a $0.241[0.035]$  rescaling of the cross section when, {e.g.}, $g_{B-L} = 0.5[0.188]$. This yields a new limit of 1.9[0.7] TeV (2.2[0.81] TeV). For reference, Tab.  \ref{tabXsec} gives $\sigma\times{\rm BR}$($Z^\prime\to ee$) for the SSM and BLSSM-IS at different $g_{B-L}$ values.

\begin{table}[!t]
\caption{{Representative $pp\to Z^\prime\to ee$ rates ($\sigma\times$ BR) for different $Z'$ masses/couplings at the LHC ($\sqrt s=8$ TeV) in the BLSSM-IS.}}
\centering
\begin{tabular}{*5l}
\hline
 $M_{Z^\prime}$ [GeV]    &  $\sigma_{\rm SSM}$ [fb]  &  \multicolumn{3}{c}{ $\sigma_{B-L}$  [fb] (with IS)}                                   \\
                               &                           &  $g_{B-L}=g_Z={e/\sin\theta_W\cos\theta_W}\ \ $ & $g_{B-L}=0.5\ \ $   & $g_{B-L}=0.8$ \\
\hline
\hline
1000                      & 170  & 6            &  41 & 105.7 \\
\hline
1500                      & 21.7  & 0.58      &  4.5 & 13.2 \\
\hline
2000                      & 3.4  & 0.087      &  0.72 & 2.3 \\
\hline
2500                      & 0.8  & 0.015      &  0.15 & 0.58 \\
\hline
3000                      & 0.21  & 0.003      &  0.04 & 0.19 \\
\hline
3500                      & 0.06  & $6\times10^{-4}$ & 0.009 & 0.06\\
\hline
\end{tabular}
\label{tabXsec}

\end{table}

\begin{figure}[h]
  \centering
  \includegraphics[height=3.5cm,width=4.5 cm,scale=0.1]{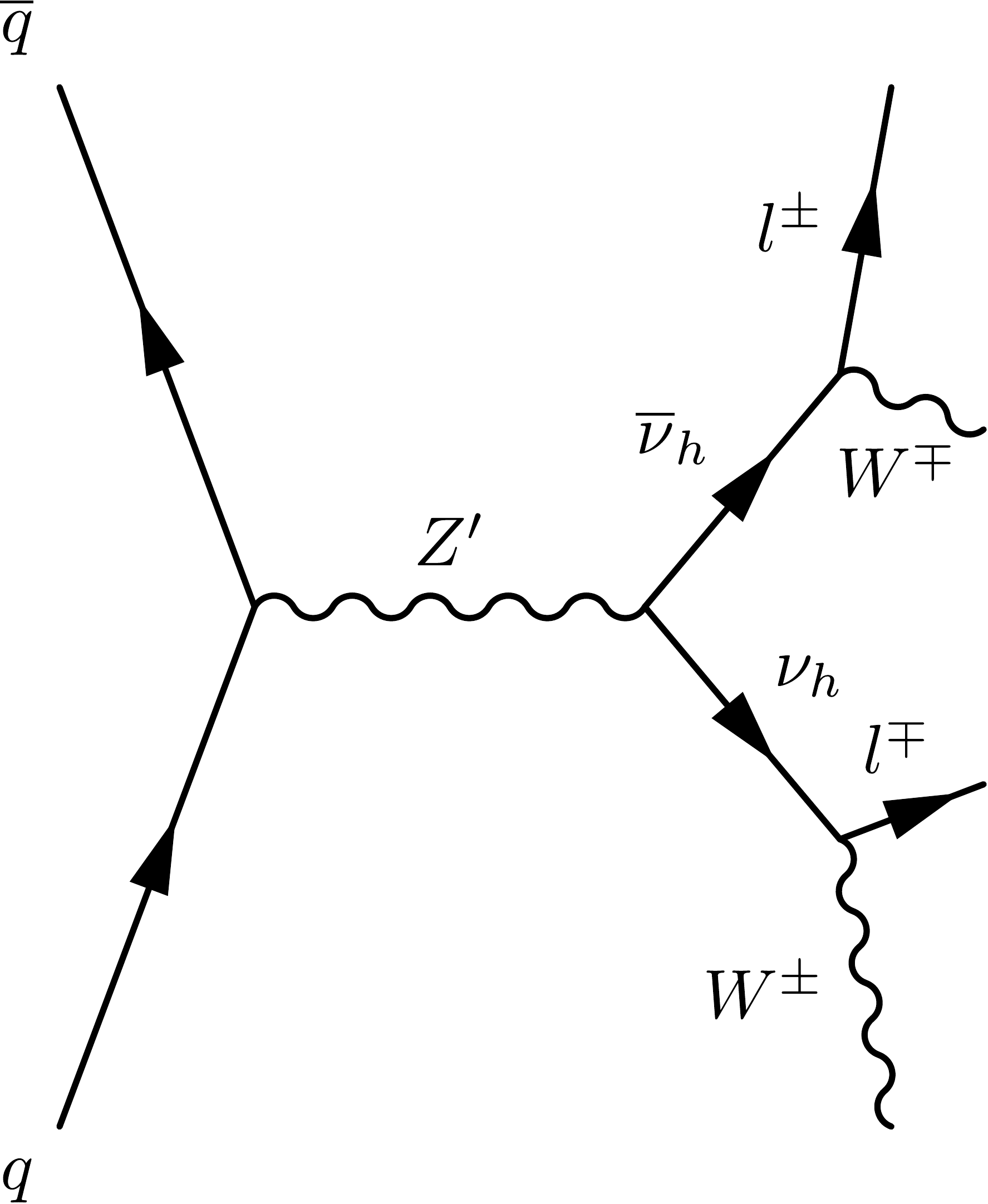}
  \caption {Feynman diagram for $ q\bar{q} \rightarrow Z^\prime \rightarrow \nu_h \bar{\nu_h} \rightarrow WW ll $.}
    \label{figFeynman}
\end{figure}

Detecting a $Z'$ signals for such small $Z'$ masses, of order 1 TeV, obtainable for $g_{B-L}\approx g_Z$, would only be circumstantial evidence for the BLSSM-IS though,
as other $Z'$ models may well feature a similar mass spectrum.  A truly smoking-gun signature of the BLSSM-IS would be to produce a $Z'$ and heavy neutrinos simultaneously. Indeed,
it turns out that the dominant production mode for heavy neutrinos at the LHC would be through the  Drell-Yan (DY)  mechanism itself,
mediated by the $Z^\prime$. The mixing between light and heavy neutrinos generates new couplings between the heavy neutrinos, the weak gauge bosons $Z,W$ and the associate leptons. These couplings are crucial for the decay of the heavy neutrinos. The main decay channel is through a $W$ gauge boson, which may decay leptonically or hadronically.
We sketch this production and decay  channel via the Feynman diagram given in Fig. \ref{figFeynman}.
Unfortunately, the very fact that $g_{B-L}$ ought to be small to comply with current LHC data which rule out $Z'$ detection in the DY induced di-lepton channel for ${\cal O}$(1 TeV) masses in turn means that such $Z'\to\nu_h\bar\nu_h$ decays also remains unaccessible. One needs a stronger $g_{B-L}$ coupling to access the latter, which thus requires higher $Z'$ masses. Hence, for the remainder of this studies, we will adopt a BLSSM-IS benchmark wherein $M_{Z'}=2.5 $ TeV.

Once such a $Z'$ state is produced and decays into $\nu_h \bar{\nu_h} \rightarrow WW ll $, one has to further sample $WW$ decays.
In case of a multi-lepton final state, one ends up with four leptons plus missing energy ($4l+ 2\nu_l$), while in case of a multi-hadronic final state states one ends up with four jets plus two leptons ($4j+2l$).
In addition, it is also possible to have a mixed final state ($2j+3l + \nu_l$).
 (Notice that one or more neutrinos would appear in the detector as missing transverse energy/momentum, $E_T^{\rm miss}$.) If two flavours of the heavy neutrinos are assumed to be degenerate in mass, one gets the same final states for the produced heavy neutrino pair with similar event rates. This will double the number of final state events but will make it difficult to distinguish between final state leptons. Therefore, throughout the current study, we consider  non-degenerate heavy neutrino masses also
including the interference between every two different flavours. (See Refs.~\cite{Basso:2008iv,Basso:2009hf,Basso:2009gg,Khalil:2006yi2,Khalil:2006yi3,Khalil:2006yi4} for alternative phenomenological
analyses in the case of the standard $B-L$ model.)

{
\begin{center}
\begin{figure}[!t]
\epsfig{file=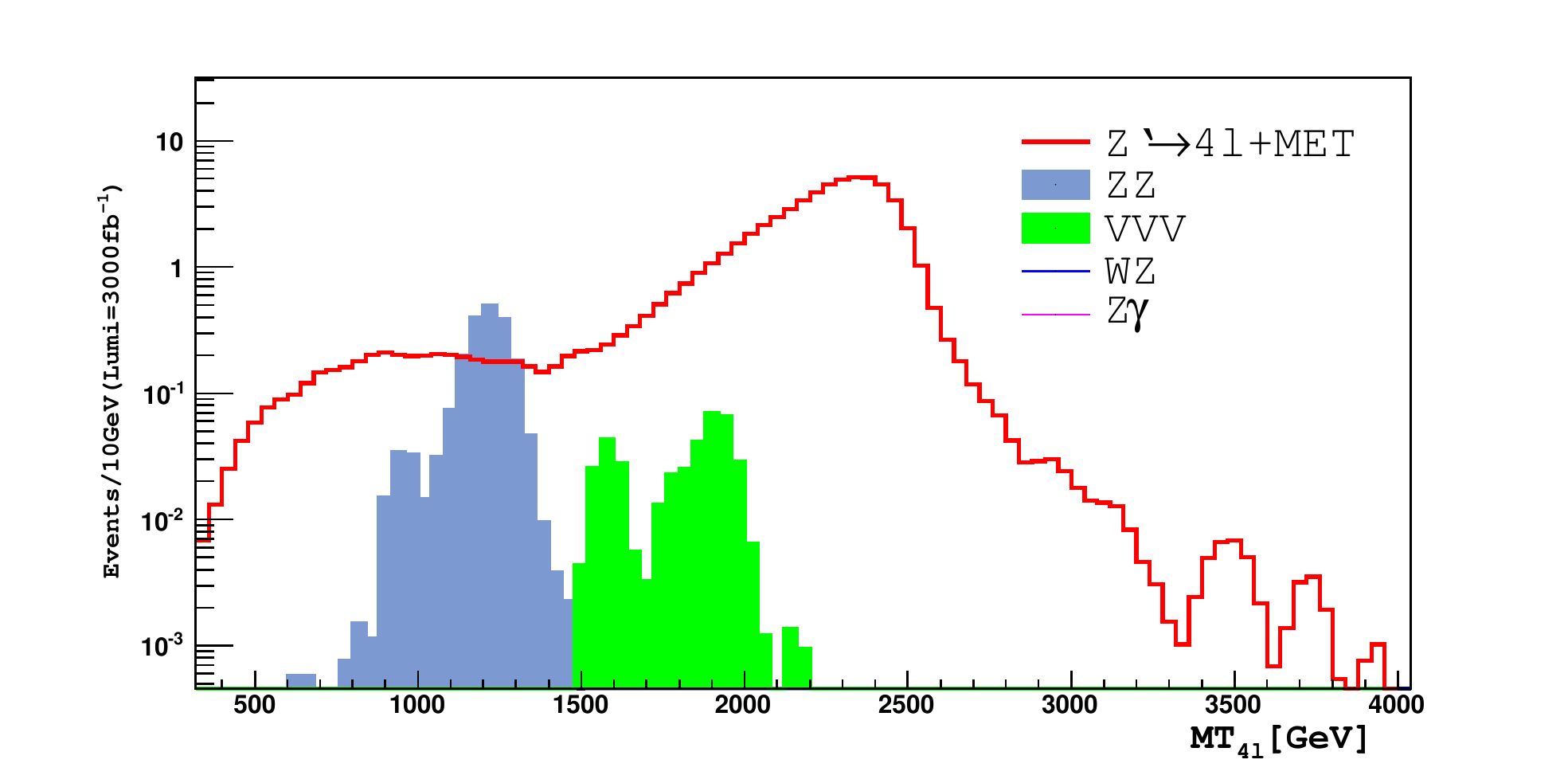, height=6.0cm,width=8.cm,angle=0}~~~ \epsfig{file=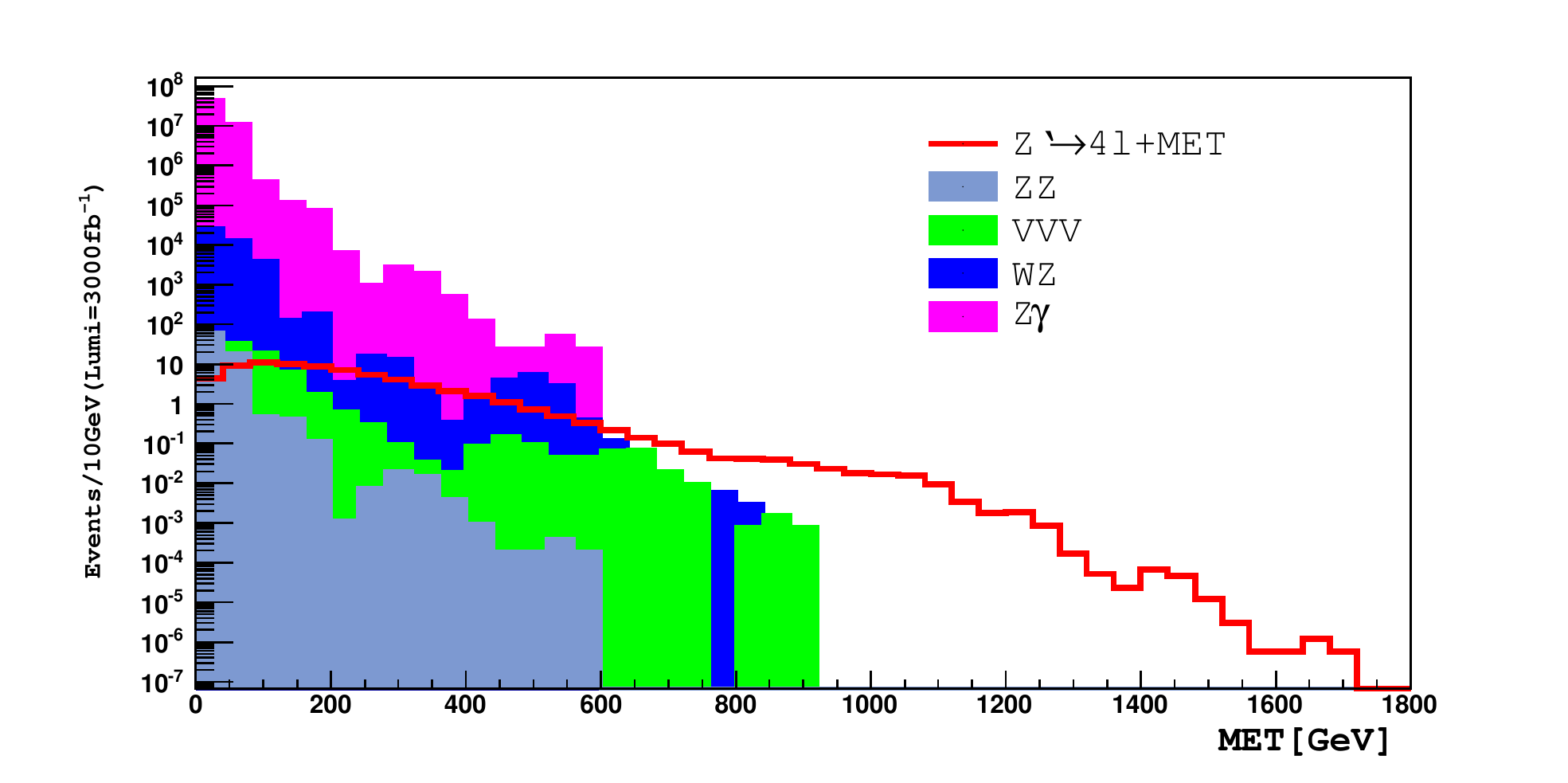,height=6.0cm,width=8.0cm,angle=0}
\caption {Number of events versus the transverse mass of the `4lepton+ $E_T^{\rm miss}$' system(left) and the missing transverse energy (right) The expected SM backgrounds are included . The luminosity assumed here is 3000 fb$^{-1}$. Note that the bin width is 10 GeV. }
 \label{fig4}
\end{figure}
\end{center}

We thus focus on the possibilities of the LHC in accessing $Z'$ decays into heavy neutrinos in the BLSSM-IS. In doing so, we have carried out full Monte Carlo (MC) event generation using
 PYTHIA \cite{Pythia} to simulate the initial and final state radiation, fragmentation and hadronisation effects. For detector effects we have used Delphes \cite{deFavereau:2013fsa}.

We consider the following benchmark: $M_{Z^\prime} = 2.5$ TeV, $M_{\nu_4}=M_{\nu_5}= 250$ GeV, $M_{\nu_6}=M_{\nu_7}= 400$ GeV and $M_{\nu_8}=M_{\nu_9}= 630$ GeV. Of the three decay signatures of the
$WW$ pair discussed
in Ref. \cite{Abdelalim:2014cxa}, i.e., $4j$, $2jl\nu$ and $2l2\nu$ ($l=e$), only the latter appeared promising, hence we only
focus here on this case, by highlighting the main results\footnote{A publication with full details and additional results is in progress \cite{progress}.}. In doing so, we produce our results at $\sqrt s=14$ TeV assuming a variable luminosity, ranging from the standard 300 fb$^{-1}$ to the tenfold increase forseen at the Super-LHC \cite{Gianotti:2002xx}. The selections
assumed in our analysis are as follows \cite{Abdelalim:2014cxa}:    a transverse momentum, $p_T$, cut of $10$ GeV and a pseudo-rapidity,  $\eta$, cut of 2 were set on each electron while the separation between two electrons, $R_{ll}$, was enforced to be 0.2.
We have assumed no restrictions on $E_T^{\rm miss}$, generated for this signature by  two neutrinos escaping detection.

 The key advantage of this channel is that it is almost background free. The main SM noise comes from $WWZ$ (three  gauge boson)  production with $\sigma(WWZ)\sim 200\  $ fb at 14 TeV \cite{Barger:1988,Hankele}. In Fig. \ref{fig4} we show the invariant mass of the `4 lepton'
system from the $Z^\prime$ signal versus the $WWZ$ background and also the transverse mass of the `4 lepton + $E_T^{\rm miss}$' system, where such a variable is defined as
\begin{equation}\label{MT}
M_T=\sqrt{(\sqrt{M^2(4l)+p_T^2(4l)}+|p_T^{\rm miss}|)^2-({\vec p}_T({4l})+{\vec p}_T^{~\rm miss})^2}.
\end{equation}
These figures indicate that the decay channel `4 lepton + $E_T^{\rm miss}$'
yields a quite clean signature and is rather  promising for probing both $Z^\prime$ and $\nu_h$ after by the end of the standard luminosity run of the LHC by using only few
simple cuts to extract the Signal ($S$) from the Background ($B$). This is clear from the achievable numbers of events left after the set of cuts mentioned above, alongside their statistical significance, $S/\sqrt{B}$,
as a function of the luminosity, see  Fig.~\ref{fig:lumi}.

\begin{center}
\begin{figure}[!t]
\epsfig{file=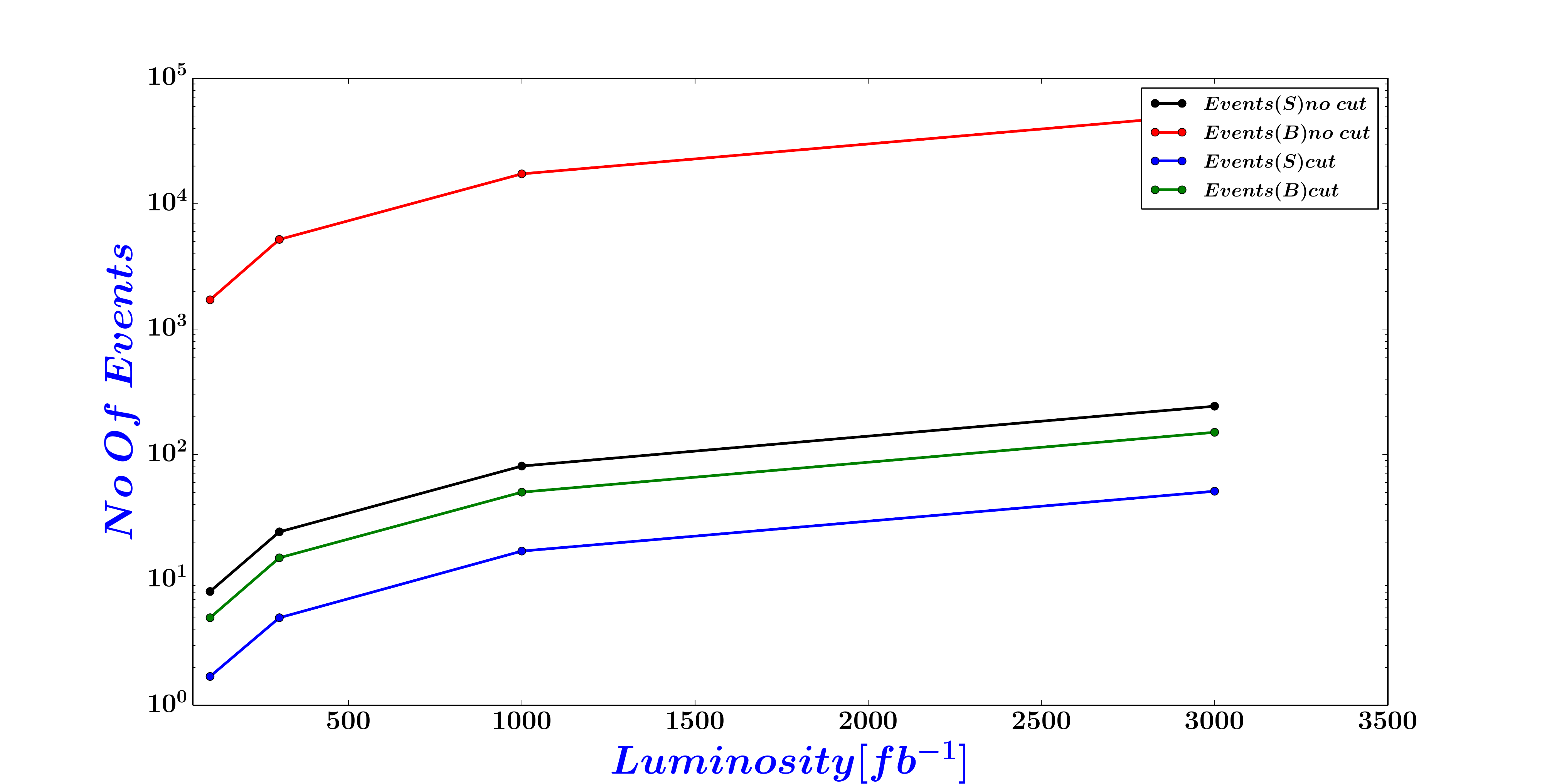, height=5.0cm,width=8.0cm,angle=0}~~~ \epsfig{file=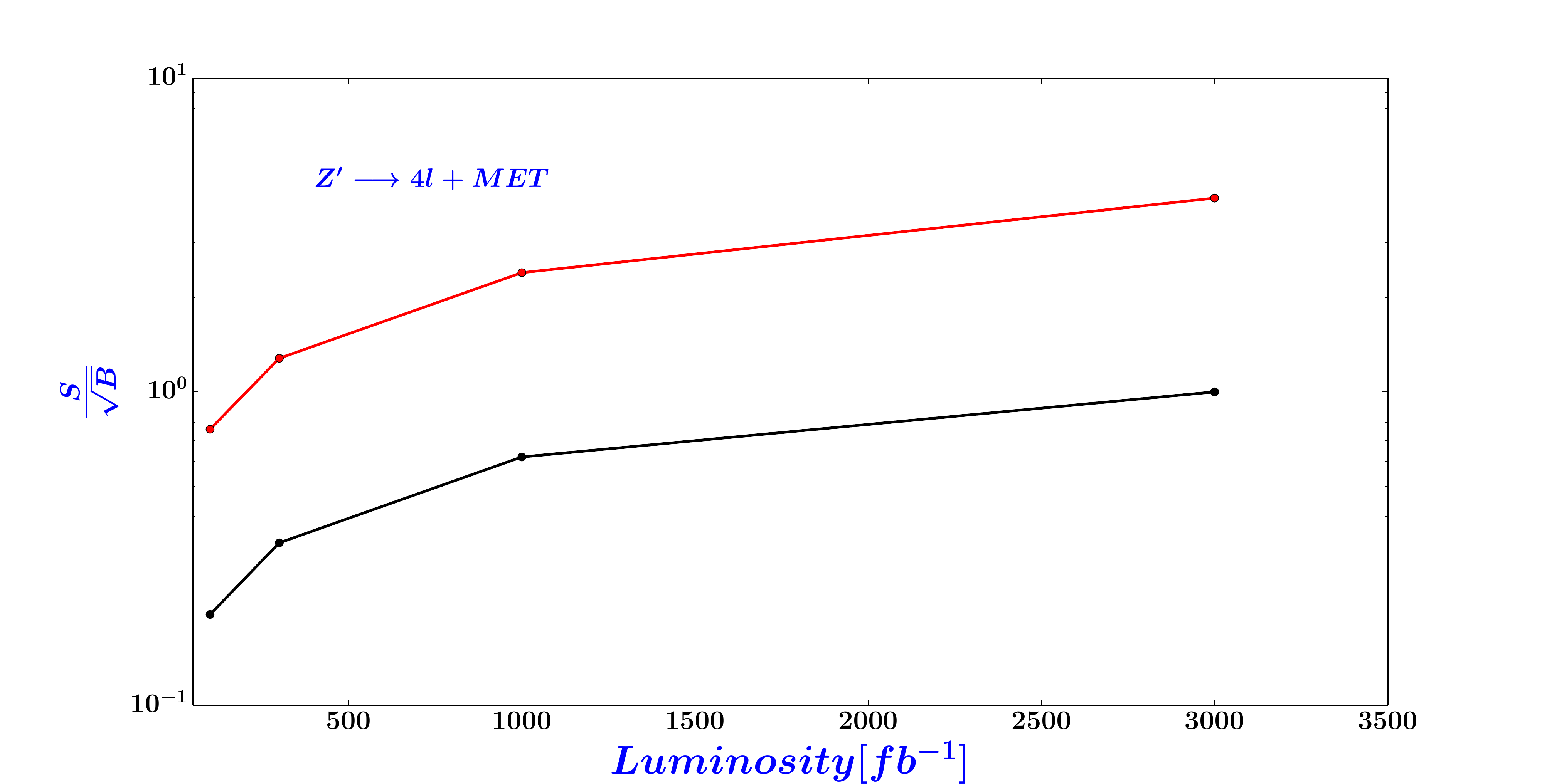,height=5.0cm,width=8.0cm,angle=0}\\
\epsfig{file=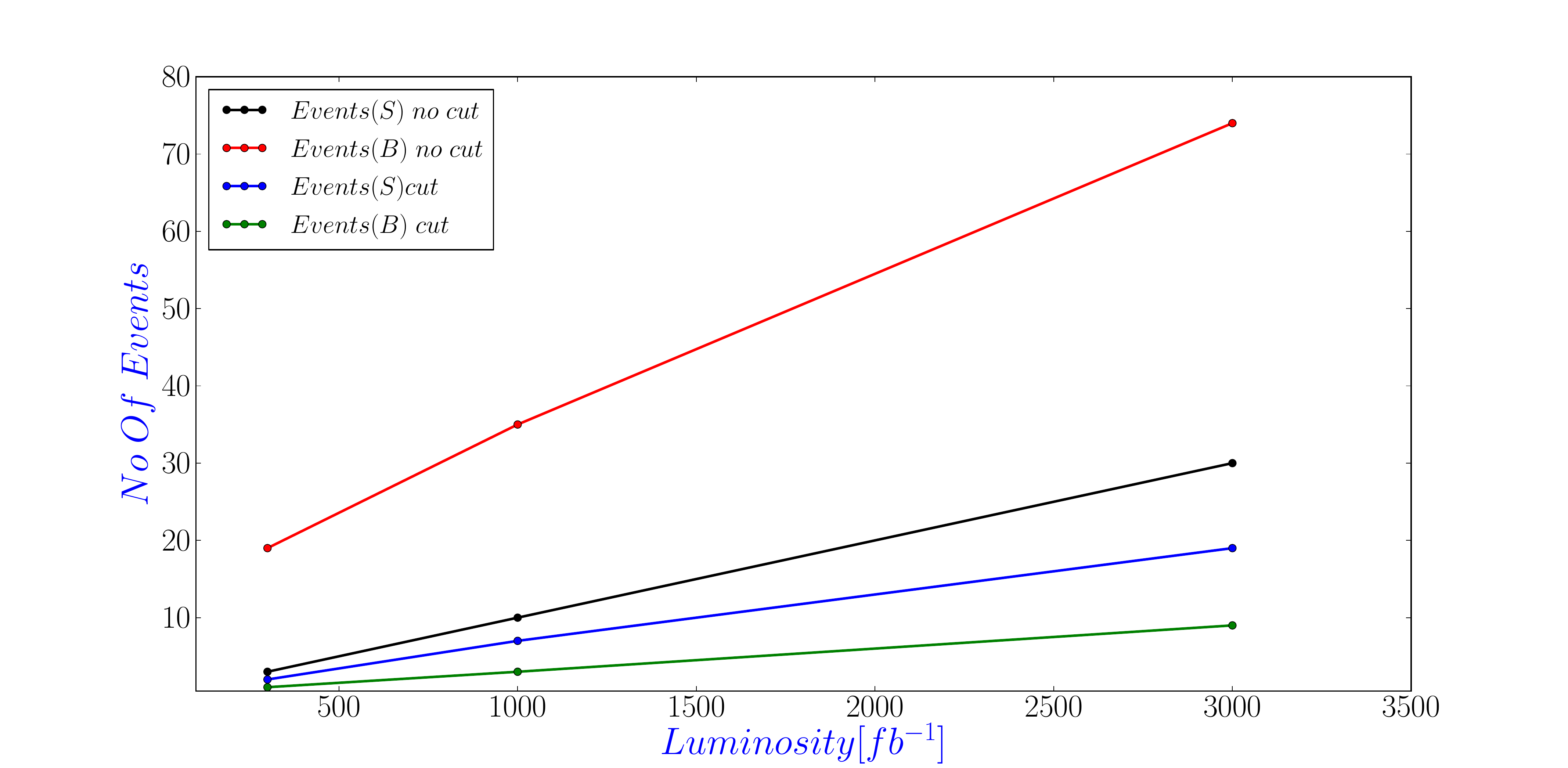, height=5.0cm,width=8.0cm,angle=0}~~~ \epsfig{file=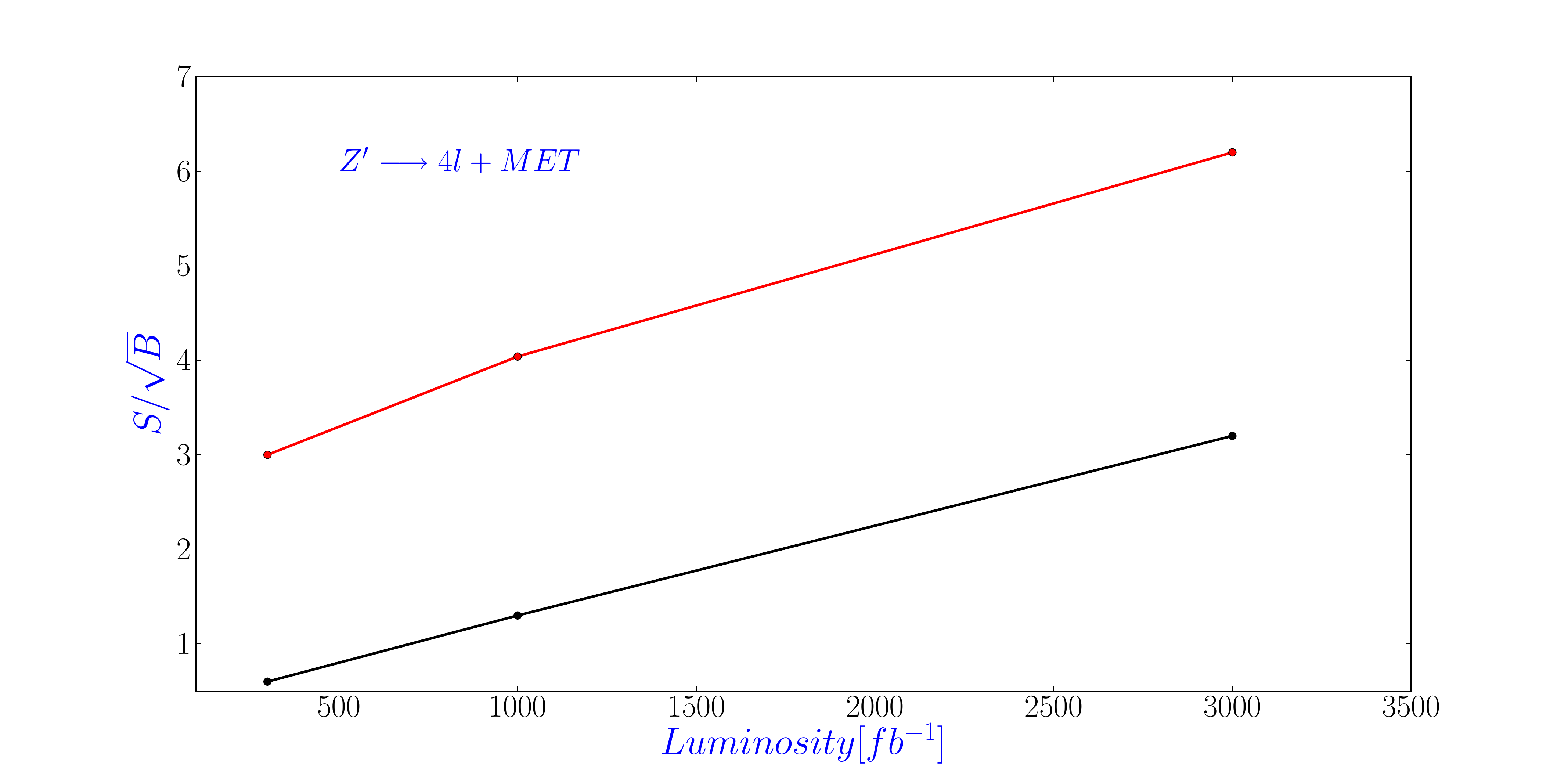,height=5.0cm,width=8.0cm,angle=0}
\caption {The $S$ and $B$ rates (left) and their significances $S/\sqrt{B}$ (right),  before and after the mass cuts,
as a function of the luminosity. At the top (bottom) the mass cut is $M({4l})>1.5$ TeV ($M_T>2$ TeV). For the significance plots, the colour scheme is as follows:
the black (red) line is before (after) the {\textcolor{blue}{corresponding}} mass cut.
}
 \label{fig:lumi}
\end{figure}
\end{center}

One thing should however be noted at this stage, concerning the statistical analysis presented in Fig.~\ref{fig:lumi}.
Herein, we have at times (i.e., depending on the luminosity)
calculated significances using the expression $ S/\sqrt{B}$ even for event numbers of $ {\cal O}(1)$,
which may not be entirely appropriate, as such an approach normally requires  large event samples  and
$B > S$.  In essence, the $S/\sqrt{B}$ method is not accurate as it is based on a $\chi^2$ distribution of test statistics which assumes that noise is negligible. Thus, as an alternative approach, we have used the `frequentest   method' too, based on Ref.~\cite{frequentist}. With reference to Sect. 2 therein, using our MC samples, we have calculated the expected significance at $\mu =0$  to reject the null, i.e., background only, hypothesis. For simplicity, we have used only the statistical error for the nuisance parameters, which can easily be extracted from our histograms.
Also, we have computed the expected significance in the signal region before and after a somewhat looser final mass cut,
in order to allow for a quantitatively sounder comparison
 (e.g., in the case of
$M_T > 1$ TeV only, as the pattern emerging from the $M(4l)$ selection is similar)\footnote{Note that the concept of  signal region is obvious  where the null hypothesis has to be rejected in the region where the signal is bumped over the background and be accepted in the control region where the sample is background dominated.}. From the plots in Fig.~\ref{fig:frequentist}, it is  clear that, at low luminosity, the difference between the frequentist and $S/\sqrt{B}$ methods is large while at higher luminosity they tend to converge, with the latter approach overestimating the significance over the former. Clearly, an appropriate
merging of the two approaches is eventually required in order to assess the feasibility of our studies at the experimental
level, depending on the actual size of the data samples. In fact, to do any cut optimisation is out of scope of this paper, as in real experimental analyses
this is actually done using Multi-Variate Analysis (MVA) techniques,  which essentially assume a set of chosen cuts for a range of values of each of these then  calculate the efficiency for each combination. Then various numerical algorithms are used to decide on the best choice of combination to finally apply in the analysis.
What we did here was instead to first characterise the signal and background based on MC truth  and then decide the relevant cuts after
plotting both of these, which we have done essentially by chosing visually the region where the former exceeds the latter. As we are unable to access such
MVA tools, which belong to the repositories of the collaborations, we cannot  credibly improve our analysis any further at this stage.

\begin{center}
\begin{figure}[!t]
\epsfig{file=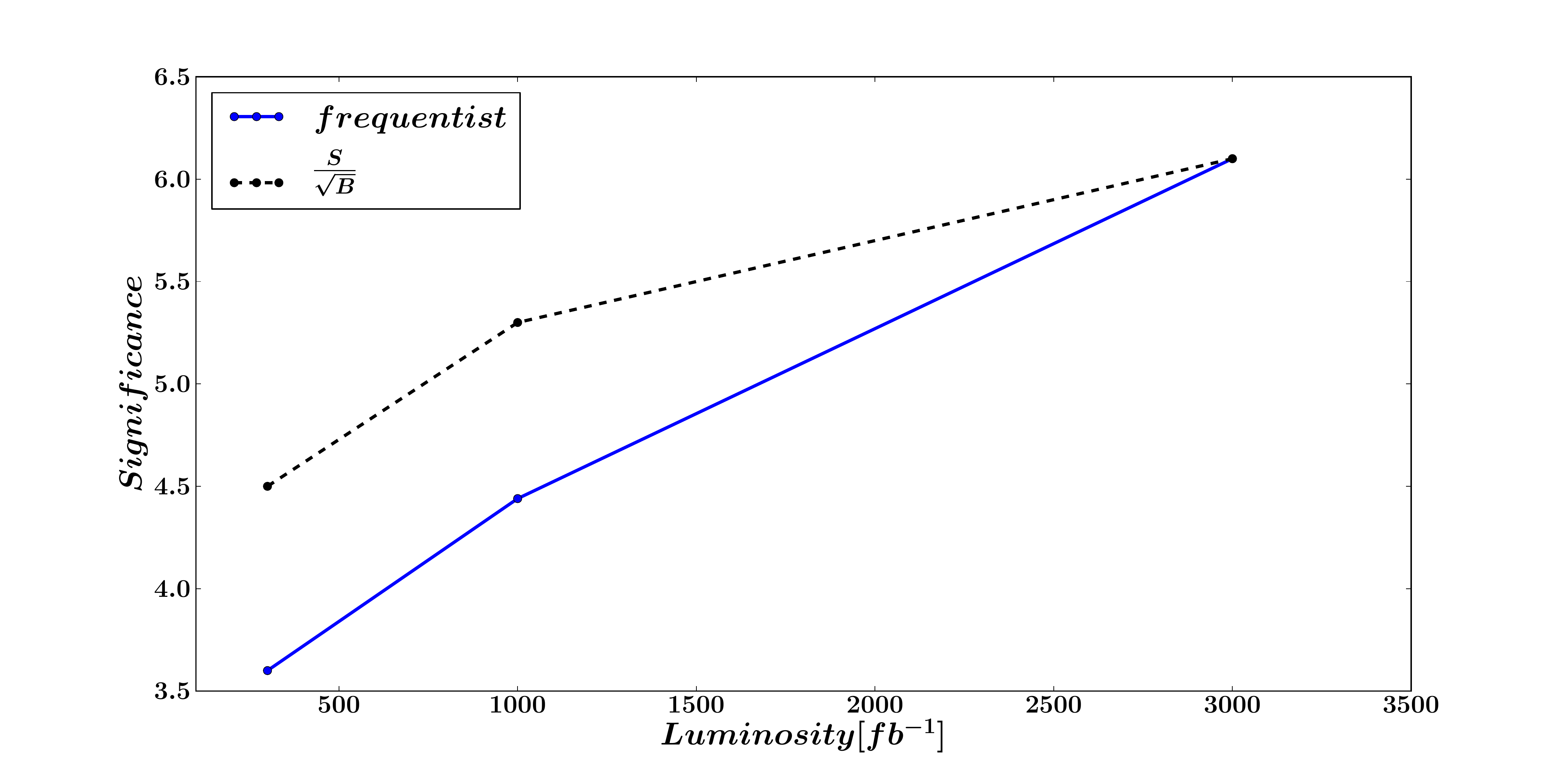, height=5.0cm,width=8.0cm,angle=0}~~~
\epsfig{file=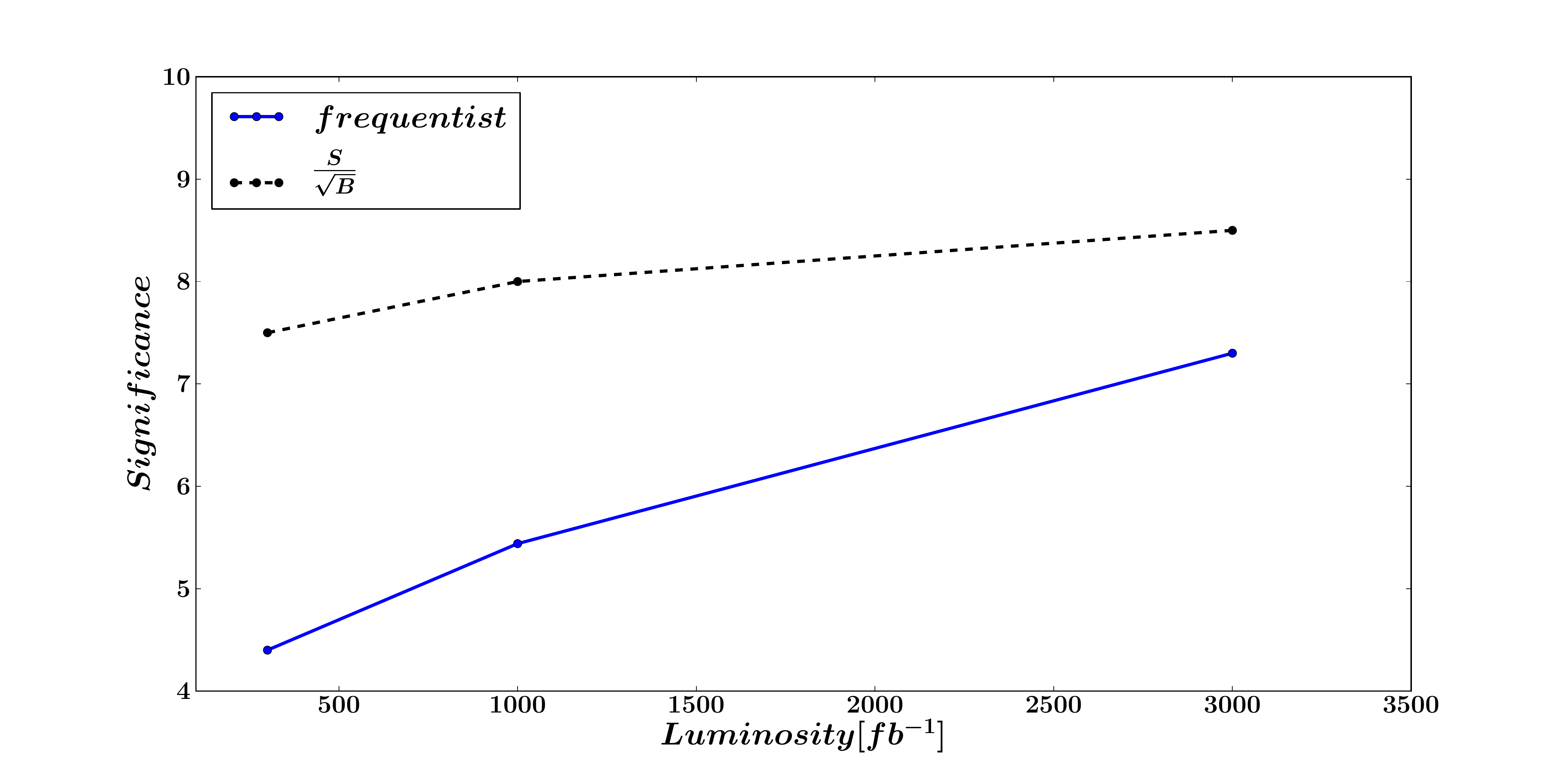,height=5.0cm,width=8.0cm,angle=0}
\caption {Comparison between the significance defined as $S/\sqrt{B}$ versus the frequentist approach described in the text before (left) and after (right) the cut  $M_T>1$ TeV.
}
 \label{fig:frequentist}
\end{figure}
\end{center}

Before closing this section, we should however comment on the influence of the SUSY spectrum on the scope of the $Z'\to \nu_h\bar\nu_h$ signal within the BLSSM. Clearly, once the assumption made so far (that all sparticles are
heavier than the $Z'$) is dismissed, the $Z'$ boson can decay via SUSY objects. This will correspond to an increased value of its total width $\Gamma^{\rm tot}_{Z'}$, which would then reflect onto the event rates of such a signal.
In fact, the latter will scale as the inverse of $\Gamma^{\rm tot}_{Z'}$. In order to quantify this (reduction) effect induced by a low mass spectrum within the BLSSM, we have revisited the benchmarks introduced in Tab. \ref{tabXsec}, excluding only the 1 TeV mass point (now ruled out by data, as discussed) and thus including the 2.5 TeV benchmark  considered so far in our MC analysis.  For each of these scenarios we have computed the ratio
\begin{equation}
R(Z')=\frac{\Gamma({Z'}\to \rm{all~BLSSM~decays})}{\Gamma({Z'}\to \rm{SM-like~decays})},
\end{equation}
the inverse of which corresponds to the rescaling factor to be applied to our event rates for the $Z'\to \nu_h\bar\nu_h$ signal in presence of a low-lying SUSY spectrum in the BLSSM. We can see  from Fig. \ref{fig:Zp-width} that $R(Z')$, obtained from the average total $Z'$ width after scanning the entire BLSSM parameter space compatible
with current experimental and theoretical constraints, is typically smaller than 3, so that the $Z'\to \nu_h\bar\nu_h$ rates will generally not be smaller than a factor 1/3 with respect to the values considered here. In fact, also recall that we have not allowed for decays into muons ($l=\mu^\pm$), which would contribute a factor of $\approx2$ towards
the signal rates in the `4 lepton + $E_T^{\rm miss}$' channel. The SM rates presented here would of course be unchanged. One should however include intrinsic SUSY backgrounds in the full analysis, as is presently being done in \cite{progress}. We can anticipate that the latter would not spoil the feasibility of such a signal as potentially (i.e., modulus experimental verification, both kinematical and statistical) established here over a sizable portion of the full BLSSM parameter space.

\begin{center}
\begin{figure}[!t]
\begin{center}
\epsfig{file=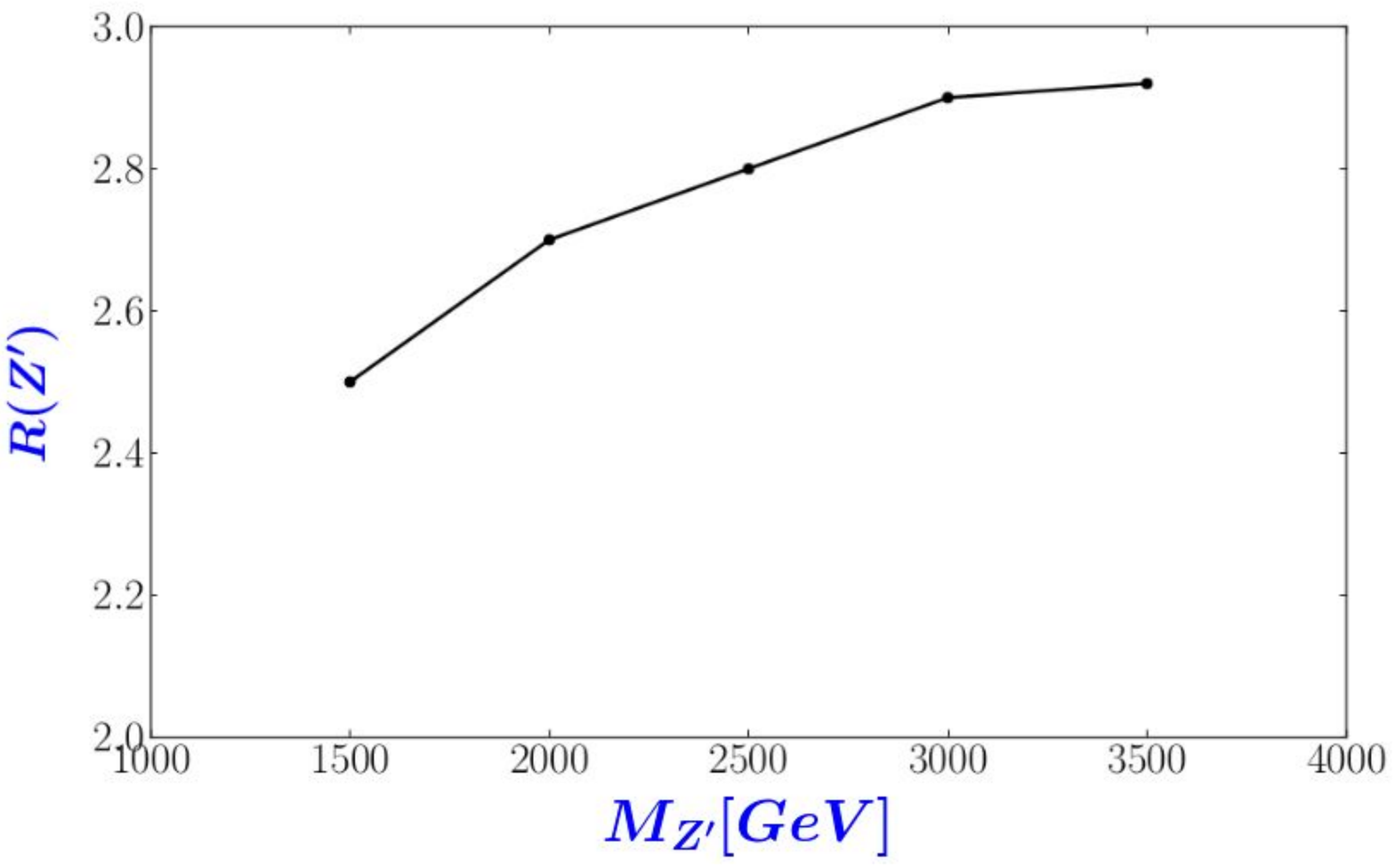, height=6.5cm,width=8.5cm,angle=0}
\end{center}
\caption {The ratio $R(Z')$ (as defined in the text) versus  the $Z'$ mass for the four heaviest benchmarks points in Tab. \ref{tabXsec}.}
 \label{fig:Zp-width}
\end{figure}
\end{center}

}

\subsection{{\large\bf Search for an Extra BLSSM-IS Higgs Boson}}
%
The Higgs decay into $ZZ \to 4l$ is one of the golden channels, with low background, to search for Higgs boson(s). The search is performed by looking for resonant peaks in the $m_{4l}$  spectrum, {\it i.e.},
the invariant mass of the $4l$ system. In CMS \cite{Chatrchyan:2013mxa}, this decay channel shows two significant peaks at 125 GeV and around/above $137$ GeV. We define by $\sigma(pp \to h')$  the total $h'$ production cross section, dominated by gluon-gluon fusion, computed for $m_{h'}=136.5$ GeV, for definiteness. (See Tab. I in
\cite{Abdallah:2014fra} for the BLSSM parameters corresponding to this specific benchmark point, which
is well compliant with current experimental limits.)  From Sect.~\ref{sect:spectrum}, it is then clear that
\be
\frac{\sigma(pp \to h')}{\sigma(pp \to h)^{\rm{SM}}} \simeq \left ( \frac{\Gamma_{{32}}}{\sin \beta}\right )^2,
\ee
(wherein the label SM identifies the SM Higgs rates computed for a 125 GeV mass),
which, for $m_{h'} \approx137$ GeV, is of order {${\cal O}(0.1)$}. Also the ratio between BRs can be estimated as
{\fontsize{10}{12}\selectfont
\be
\frac{\rm{BR}( h'\to ZZ)}{\rm{BR}(h\to ZZ)^{\rm{SM}}} \simeq \left(1+\frac{\Gamma^{\rm{SM}}_{h\rightarrow W W^*}}{\Gamma^{\rm{SM}}_{h\rightarrow b \bar{b}}}\right)\frac{F(M_Z/m_{h'})}{F(M_Z/m_{h})^{\rm SM}}
 \times\left[\left(\frac{\Gamma_{31}\sec{\beta}}{\Gamma_{32} \sin{\beta}+ \Gamma_{31} \cos{\beta}}\right)^2+2F\left(\frac{M_W}{m_{h'}}\right)\right]^{-1},~
\ee}
where
$$ F(x)=\frac{3(1-8 x^2+20 x^4)}{(4 x^2-1)^{1/2}}\arccos\left(\frac{3 x^2-1}{2 x^{3}}\right)$$
 \begin{equation}-\frac{1-x^2}{2 x^2}(2-13 x^2+47 x^4)-\frac{3}{2}(1-6 x^2 +4 x^4)\log{x^2}.\end{equation}
\begin{figure}[t]
\begin{center}
\epsfig{file=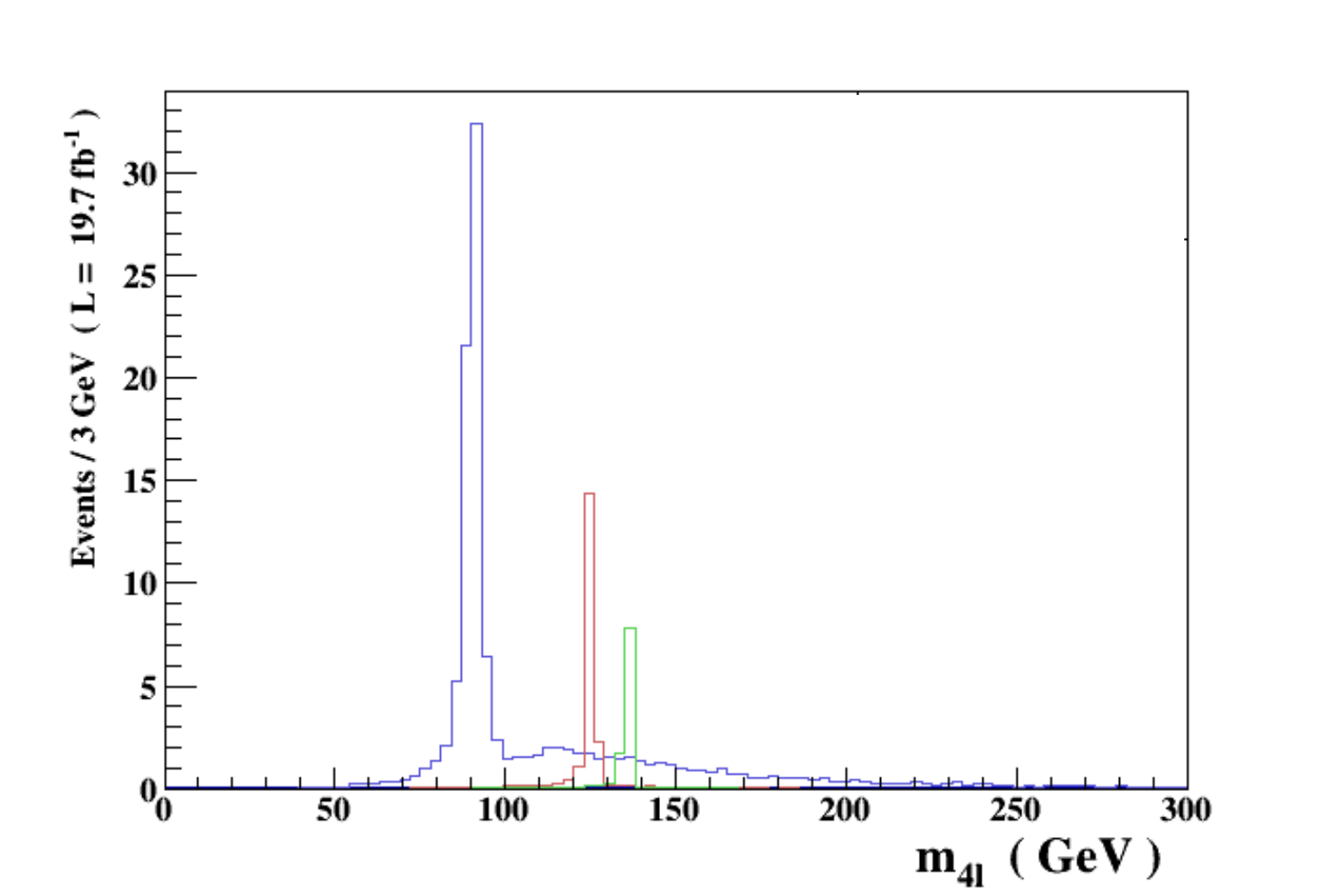,height=5cm,width=8cm,angle=0}
\caption {The number of events of the processes $pp \to Z\to 2l \gamma^* \to 4l$ (blue), $pp \to h \to ZZ \to 4l$ (red) and $pp \to h' \to ZZ \to 4l$ (green) versus the invariant mass of the out going particles (4-leptons), ${m}_{4l}$.}
\label{fig2}
\end{center}
\end{figure}
First  we analyse the kinematic search for the BLSSM-IS Higgs boson, $h'$, in the decay channel to $ ZZ \to 4l$.
In Fig.~\ref{fig2}, we show the invariant mass of the 4-lepton final state from $ pp \to h' \to ZZ \to 4l$ at $\sqrt{s}={8}$ TeV, after applying a $p_T[|\eta|]$ cut of 5[2.5] GeV on each of the four leptons \cite{Abdallah:2014fra}\footnote{Transverse momentum and pseudo-rapidity thresholds are slightly different for
electrons and muons, but accounting for this subtlety would not change our conclusions.}. The SM model backgrounds from the $Z$ and {125} GeV Higgs boson decays,  $pp \to Z\to 2l \gamma^* \to 4 l$ and $pp \to h \to ZZ \to 4l$, respectively, are taken into account, as
demonstrated by the first two peaks in the plot (with the same $p_T$ requirement). It is clear that the third peak at ${m}_{4l} \sim 137$ GeV, produced by the decay of the BLSSM-IS Higgs boson $h'$ into $ZZ \to 4 l$, can reasonably well account for the events observed by CMS \cite{Chatrchyan:2013mxa} with the 8 TeV data. This is shown in Tab.  \ref{tab2}, where the mass interval in m$_{4l}$ that we have investigated to extract the $h'$ signal
is wide enough to also capture another prominent 145 GeV anomaly seen in the
same CMS data set.
(The bin width used in   \cite{Chatrchyan:2013mxa} is 3 GeV so potentially able to separate the two peaks at 137 and 145 GeV, yet combining the handful of events under the two peaks makes statistical sense given, on the one hand, the small mass difference and, on the other hand, the fact that for
masses so close to the $WW$ threshold one can find intrinsic BLSSM  Higgs widths of order GeV.)
\begin{table}[t]
\begin{center}
\begin{tabular}{|c|c|c|c|c|}
\hline
\multicolumn{5}{|c|}{ Number of events for $ 19.7\; \rm{fb}^{-1}$ at $\sqrt{s}=8$ TeV}\\
\hline
{Higgs mass}&Observed&Expected&\multicolumn{2}{c|}{Background}\\
\cline{4-5}
& (CMS)&(BLSSM) &$Z\to 2l\gamma^*$&$h \to ZZ$\\
\hline
125 GeV & 25 & 18.5&6.6&-\\
\hline
136.5 GeV & 29 & 10.2&9.15&0.8\\
\hline
\end{tabular}
\caption{The observed (by CMS) and expected (from the BLSSM) number of events in a mass window around $m_h=$ 125 GeV ($121$ GeV $<m_{4l}< 131 $ GeV) and $m_{h'}=136.5$ GeV ($ 131$ GeV $<m_{4l}< 152 $ GeV) in the $ZZ\to 4l$ channel compared to the expected (dominant) $pp\to Z \to 2l\gamma^*\to 4l$ and $pp \to h \to ZZ \to 4l$ backgrounds.}
\label{tab2}
\end{center}	
\end{table}

Next we turn to the di-photon channel, which provides the greatest sensitivity for Higgs boson discovery in the
intermediate mass range ({i.e.}, for Higgs masses below $2M_W$)\footnote{The effects of light SUSY particles leading to a possible enhancement of the
di-photon signal strength of the SM-like Higgs boson were studied in \cite{diphotons}.}. Like the SM-like Higgs, the $h'$ decays into two photons through a triangle-loop diagram dominated by (primarily) $W$ and (in part) top quark exchanges.  As shown in Sect.~\ref{sect:spectrum}, the couplings of the $h'$ with top quarks and $W$ gauge bosons are proportional to some combinations of $\Gamma_{{31}}$ and $\Gamma_{{32}}$, which may then lead to some suppression or enhancement in the partial width $\Gamma(h' \to  \gamma \gamma)$. In the SM,  ${\rm{BR}}(h\to \gamma \gamma) \simeq 2\times 10^{-3}$. Similarly, in the BLSSM, we have found that, for our $m_{h'}= 136.5$ GeV benchmark, the BR of $h'$ in photons amounts to {$2.15 \times 10^{-3}$}, hence within current experimental constraints.
 \begin{figure}[t]
 \begin{center}
\epsfig{file=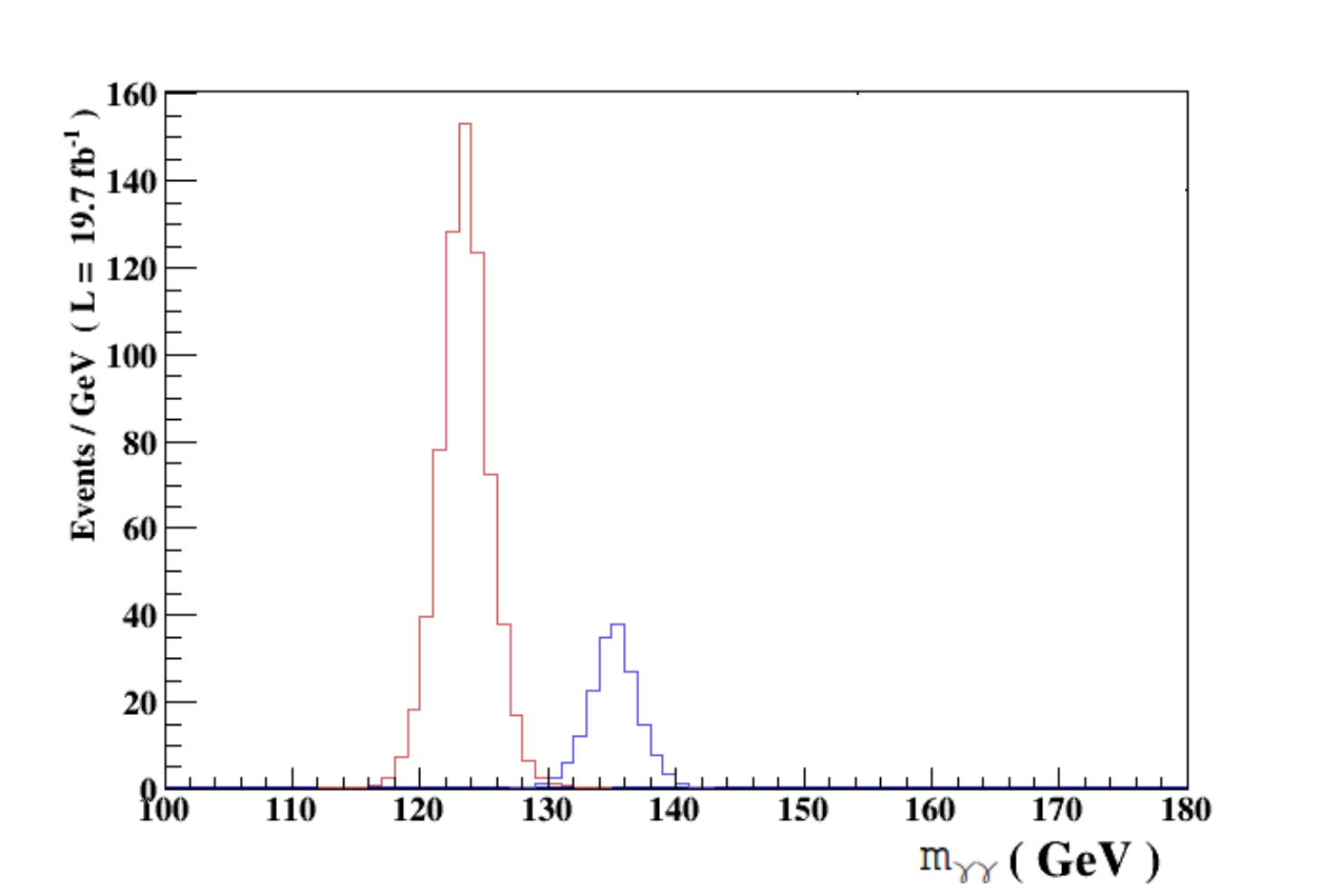,height=5cm,width=8cm,angle=0}
\caption { The number of events of the processes $pp \to h \to \gamma \gamma$ (red), $pp \to h' \to \gamma \gamma$ (blue) versus the invariant mass of the outgoing particles (di-photons), $m_{\gamma\gamma}$.}
\label{fig3}
\end{center}
\end{figure}

\begin{table}[t]
\begin{center}
\begin{tabular}{|c|c|c|}
\hline
\multicolumn{3}{|c|}{Number of events for $19.7\; \rm{fb}^{-1}$ at $\sqrt{s}=8$ TeV}\\
\hline
{Higgs mass}& Observed (CMS) & Expected (BLSSM)\\
\hline
$125$ GeV & 610 & 666\\
\hline
$136.5$ GeV & 170 &177 \\
\hline
\end{tabular}
\caption{The observed (by CMS) and expected (from the BLSSM) number of events (after subtracting background) in a mass window around $m_h=$ 125 GeV ($120$ GeV $<{m_{\gamma\gamma}}< 130 $ GeV) and $m_{h'}=136.5$ GeV ($ 131$ GeV $<{m_{\gamma\gamma}}< 141 $ GeV) in the $\gamma\gamma$ channel.}
\label{tab3}
\end{center}
\end{table}
The distribution of the di-photon invariant mass is presented in Fig.~\ref{fig3} for a centre-of-mass energy $\sqrt{s}={8}$ TeV \cite{Abdallah:2014fra}. Again, here, the observed $h\to \gamma\gamma$ SM-like signal around 125 GeV is taken as background while the $Z\to \gamma\gamma$ background can now be ignored \cite{Moretti:2014rka}. As expected, the sensitivity to the $h'$ Higgs boson is severely reduced with respect to the presence of the already observed Higgs boson, yet a peak is clearly seen at 136.5 GeV and is very compatible with the excess seen by CMS \cite{CMS:2013wda}.
 This is shown in Tab.  \ref{tab3}.
 In fact, ATLAS results point in the same direction as well \cite{ATLAS-CONF-2013-012,Aad:2014eha}, see \cite{Khalil:2015vpa} and references therein. For example, in
\cite{ATLAS-CONF-2013-012},
they have a slightly worse resolution, of 2 GeV, and use a mixed  sample of 7 TeV (with 4.8 fb$^{-1}$) and 8 TeV (20.7 fb$^{-1}$) data. If we integrate five bins around 125(137) GeV, see Fig. 3 of \cite{ATLAS-CONF-2013-012}, we obtain
approximately 700(145) events above the background.
 It is worth mentioning that here we consider both the gluon-gluon fusion and  vector-boson fusion modes for both $h$ and $h'$ production.

Before closing this section, we should also mention that the $h'\to\gamma\gamma$ enhancement found in the BLSSM-IS may be mirrored in the $\gamma Z$ decay channel \cite{Hammad:2015eca}
for which, at present, there exists some constraints, albeit not as severe as
in the $\gamma\gamma$ case. We can anticipate  (see \cite{Hammad:2015eca}) that the BLSSM-IS regions of parameter space studied
here are consistent with all available data.

\section{Conclusions\label{sect:summary}}

In summary, in this mini-review, we have introduced the reader to the minimal SUSY version of the well established
$B-L$ model with an IS mechanism, that we have termed as BLSSM-IS. This scenario nicely combines the theoretically
appealing features of SUSY with  key experimental evidence of Beyond the SM (BSM)
physics in the form of neutrino masses.

Initially, we  have proceeded with the construction of the BLSSM-IS Lagrangian, followed by an illustration of how dynamical
EWSB naturally
occurs via RGE evolution
starting from an mSugra inspired model configuration at high scales.  Then, we have described the emerging
particle spectrum, by singling out the dynamics in the three specifically BLSSM-IS sectors: {i.e.}, the $Z'$, Higgs and (s)neutrino parts. In three separate subsections we have in fact derived the relevant masses and couplings.

As EWSB and $B-L$ breaking both occur close to the SUSY mass scale of order 1 TeV, the BLSSM-IS also bears interesting
phenomenological manifestations at the LHC in the three aforementioned sectors. Therefore, we have studied next the hallmark signals of this scenario in turn. Firstly, we described $Z'$ production and decay into a variety of leptonic and hadronic signatures proceding via heavy neutrinos\footnote{The case of sneutrino mediated channels has been tackled in
Ref.~\cite{Abdallah:2015hma}.}, all leading to detectable signals at Run 2 of the CERN machine. Secondly, we highlighted the striking feature of the BLSSM-IS in the Higgs sector, in the form of a possible additional light Higgs resonance yielding sizable $\gamma\gamma$ and $ZZ$ decays which may even explain some anomalies around a mass of 137 to 145 GeV
present already in the ATLAS and CMS data of Run 1 of the LHC and which have not yet been ruled out by current Run 2
samples (neither $\gamma\gamma$ \cite{Run2-AA} nor $ZZ$ \cite{Run2-ZZ} ones).

In short, the BLSSM-IS represents a viable realisation of SUSY, compliant with all current data and giving distinctive signatures at the LHC which will enable one to disentangle it from alternative BSM scenarios. These include
the simultaneous production of a $Z'$ and heavy (s)neutrinos. Further, these can  possibly be accompanied by a second light CP-even Higgs boson signal (precluded to the MSSM), but not by a 750 GeV one \cite{750GeV-ATLAS,CMS:2015dxe}, as the BLSSM-IS specific CP-even
 Higgs state is heavier than its MSSM counterpart
  (see \cite{Hammad:2016trm} for very recent results in this respect).

\section*{Acknowledgements}
The work of SK is partially supported by the ICTP grant AC-80.
SM is supported in part through the NExT Institute. The work of SK and SM is also funded through the
grant H2020-MSCA-RISE-2014 no. 645722 (NonMinimalHiggs). We thank Ahmed Hammad and Juri Fiaschi for assistance in preparing this manuscript.


\end{document}